%% file: resummation.tex
\documentclass[letterpaper,11pt]{article}
\pdfoutput=1
\usepackage{jheppub}
\usepackage{amsfonts,amsmath,amssymb}
\usepackage{graphicx,caption,subcaption}
\usepackage{resummation_macros}
\input{resummation_figures}

\input{resummation_tables}

\title{\vspace{4cm} Resummation and S-duality in N=4 SYM}
\author{Christopher Beem,\!$^a$ Leonardo Rastelli,\!$^b$
Ashoke Sen,\!$^{c,d}$ and Balt C. van Rees$^b$ \\
~\\
$^a${\it Simons Center for Geometry and Physics, SUNY, Stony Brook, NY 11794-3636, USA \\
$^b$C. N. Yang Institute for Theoretical Physics, SUNY, Stony Brook, NY 11794-3840, USA \\
$^c$Harish-Chandra Research Institute, Chhatnag Road, Jhusi, Allahabad 211019, India\\
$^d$School of Physics, Korea Institute for Advanced Study, Seoul 130-722, Korea}
}

\abstract{%
We consider the problem of resumming the perturbative expansions for anomalous dimensions of low twist, non-BPS operators in four dimensional $\NN=4$ supersymmetric Yang-Mills theories. The requirement of S-duality invariance imposes considerable restrictions on any such resummation. We introduce several prescriptions that produce interpolating functions on the upper half plane that are compatible with a subgroup of the full duality group. These lead to predictions for the anomalous dimensions at all points in the fundamental domain of the complex gauge coupling, and in particular at the duality-invariant values $\tau=i$ and $\tau=\exp(i\pi/3)$. For low-rank gauge groups, the predictions are compatible with the bounds derived by conformal bootstrap methods for these anomalous dimensions; within numerical errors, they are in good agreement with the conjecture that said bounds are saturated at a duality-invariant point. We also find that the
anomalous dimensions of the lowest twist operators lie within an extremely narrow window
around a straight line as we vary the moduli of the theory over the two dimensional 
fundamental domain. 
}

\preprint{YITP-SB-13-17, HRI/ST/1303}  
\notoc

\begin{document}

\maketitle

\vfill\eject

\flushbottom

\tableofcontents
\section{Introduction}\label{s1}

The last fifteen years have brought forth enormous progress in our understanding of four-dimensional $\NN=4$ supersymmetric Yang-Mills (SYM) theories in the planar limit. Although many of these developments were stimulated by the discovery of AdS/CFT duality,
there are by now a large number of computational techniques available directly in field theory. Nevertheless, investigations of non-planar physics beyond perturbation theory or supersymmetric observables remain in their infancy.

Recently, a new class of results  for these theories has been obtained in \cite{1304.1803} via conformal bootstrap methods, as pioneered in \cite{BootstrapMethods}. Rigorous bounds were derived for the anomalous dimensions of leading twist operators of various spins appearing in the operator product expansion (OPE) of a single four-point function. The bounds depend solely on the central charge of the theory -- they are independent of the complexified gauge coupling -- and they constitute truly non-perturbative results for the theory. No planar approximation is required.

The generality of the results of \cite{1304.1803} obfuscates more detailed properties of the observables in question, namely the variation of the anomalous dimensions over the conformal manifold parametrized by the coupling $\tau= \frac{\theta}{2\pi} +\frac{4\pi\,i}{\gym^2}$. At weak coupling the answer is known perturbatively, and by S-duality the result at strong coupling is also known. The problem of interest is to understand what happens at finite coupling. In this paper, we pursue an approximate answer to this question by looking for simple functions that smoothly interpolate between weak and strong coupling limites. A similar approach has been implemented recently to study the mass of the stable, non-BPS state in heterotic/type I string theory \cite{1304.0458}.

We make use of several different interpolating functions, which we review in \S\ref{s2}. We describe their application to the anomalous dimensions of local operators in $\NN=4$ SYM in \S\ref{s3}. This ultimately leaves us with several interpolating functions of the Yang-Mills coupling and theta angle that are guaranteed to reproduce the correct perturbative results in the weak-coupling limit. In general, we find good agreement between these different interpolating functions in their domains of mutual validity; this includes interpolations defined to match the perturbative answer to different orders. For example, for the Konishi anomalous dimension in the $SU(2)$ theory we find that with $\theta=0$ the variation between methods amounts to less than 15\% of the mean. When accounting for the tree-level contribution, this amounts to about 5\% error in the actual scaling dimension. For $\theta=1/2$, the variation is slightly larger -- 20\% of the mean for the anomalous dimension, corresponding to 7.5\% for the full scaling dimension. It is natural to expect the results to become worse for gauge groups of larger dimension. The effective coupling $\gym^2N$ then takes larger values around the duality symmetric point ($\gym \sim 3.5$), so it is less likely that perturbative results together with duality will be sufficient to accurately determine the behavior of the function.

A consistency check for our results is that, subject to the aforementioned uncertainties, they are compatible with the bounds derived in \cite{1304.1803}. In this context, the relevant question is: \emph{are the bounds saturated at some point on the conformal manifold?} In \cite{1304.1803} this was conjectured to be the case at one of the orbifold points, $\tau=i$ or $\tau=\exp(i\pi/3)$. For the benefit of the reader, we reproduce here the results of our resummations for these values of $\tau$, along with the upper bound and a `corner' value derived for these quantities in \cite{1304.1803}. As we explain further in \S\ref{s5}, the corner value is a natural best estimate for these operator dimensions based on numerical bootstrap results.

\spinZeroTwoTableNoCaption

The uncertainty attached to these results is large enough to prevent us from making a definitive statement, but the values of the anomalous dimension at $\tau=i$ and $\tau=\exp(i\pi/3)$ are sufficiently close to the bounds to be suggestive of bound saturation at one of these points. This is especially so given that the bounds are expected to lower somewhat upon improving the numerics used in the methods of \cite{1304.1803}. We have used perturbative results up to four loops to arrive at this result. The error estimates are conservative; we observe that the four loop results for all of the interpolating functions lie between two and three loop results, and hence take the mean two loop result and the mean three loop results as the allowed range for any given quantity. A five loop result (including non-planar corrections) would be likely to improve the situation.

Another interesting point to explore is the image of the conformal manifold in the space of dimensions of the lowest twist operators. Since the conformal manifold is two dimensional, we expect that the allowed values of the anomalous dimensions will trace out a  two dimensional submanifold in the space of anomalous dimensions as we vary $\tau$ over the fundamental domain. The projection of this submanifold to any two dimensional space labelled by a pair of  anomalous dimensions is also expected to be two dimensional. We use our interpolating functions to identify this submanifold in the space of anomalous
dimensions of lowest twist spin zero and spin two operators, and encounter a surprise: within the accuracy of our plots, the submanifold is a one dimensional straight line instead of a two dimensional subspace. This result appears in Fig.~\ref{fconformal}. Of course we do not expect this to be an exact result -- 
  a constant slope is inconsistent with perturbation theory, and we know that the subspace must acquire a finite width from 
the non-perturbative $\theta$ dependence -- but what our analysis shows is that the allowed values of the anomalous dimensions lie within
a very narrow band around this straight line. The maximum deviation of our interpolating function from this straight line is about $0.6\%$ over the entire range. Furthermore, different approximations and different loop orders all lead to the same result, suggesting that this result is much more robust and accurate compared to the actual value of the anomalous dimension at a given value of $\tau$.

We present all of these results in \S\ref{s4}, and make some final comments in \S\ref{s5}. Appendix \ref{sa} contains the interpolation formul\ae\ that we use for our analysis.

\section{Symmetric interpolating functions}\label{s2}

Before defining our interpolating functions, let us briefly provide some context for the approach employed in the present note. There exist a variety of sophisticated techniques for resumming perturbative expansions in quantum mechanics and quantum field theory (see, \eg, chapter 16 of \cite{KleinertBook}). In particular, when the series in question is Borel summable, powerful techniques can be brought to bear upon the problem. Interestingly, it has recently been argued that extended supersymmetric quantum field theories in four dimensions are always Borel summable \cite{1206.1890}. Then, with some additional knowledge of the large order behavior of perturbation theory, one can gain a great deal of insight into the behavior of the function in question at finite values of the coupling.

Despite the utility of integral transformations such as the Borel transform in resummation procedures, they make it difficult to impose upon the function of interest symmetries such as those implied by duality in $\NN=4$ SYM. It would be interesting to find appropriate integral transforms that tame the asymptotic behavior of perturbation theory while being compatible with duality, so as to combine the resulting symmetry constraints with a more detailed consideration of the analytic properties of the anomalous dimensions. For the present purposes, though, these dualities represent one of the most powerful pieces of information available to us, so we perform our interpolations directly at the level of the anomalous dimensions. The underlying assumption driving the present work is the following:

\vspace{12pt}
\noindent\emph{For low rank gauge groups, the effective coupling at the most strongly coupled points on the conformal manifold is not very large. 
Furthermore, the strength of non-perturbative corrections is controlled by the factor $\exp(-2\pi \, \text{\rm Im}\, \tau)$, which takes values of $.002$ and $.004$ at $\tau=i$ and $\tau=\exp(i\pi/3)$, respectively. Consequently, the anomalous dimensions should be well approximated by simple functions with the correct duality properties and perturbative expansions.}
\vspace{0pt}

Roughly speaking, our strategy is to construct interpolating functions that are as simple as possible while being invariant under some symmetries imposed by S-duality of $\NN=4$ SYM. The actual anomalous dimensions will be invariant under the action of the full PSL$(2,\mathbb{Z})$ modular group, so it may seem that we should search for interpolating functions that are modular invariant. However, such functions generally suffer from a certain amount of ambiguity: it is not clear how to define the ``simplest'' modular-invariant functions, and the results may depend substantially on the choices that are made. 

We instead choose to impose a lesser degree of symmetry on the problem by finding interpolating functions that are invariant under a \emph{finite-order subgroup} of the full modular group. By construction, these interpolating functions are most accurate at weak coupling where their series expansions are matched. In the strongly coupled region -- say near a fixed point of one of these finite order subgroups -- we expect the most accurate result to come from the interpolating function which is invariant under the corresponding symmetry. As we move away from this fixed point,  other symmetries that have not 
been taken into account will become relevant and the results should become worse. As we discuss in \S\ref{s4}, this means that we must exercise some care in interpreting our results. First, though, we describe our prescriptions for creating interpolating functions that are invariant under finite-order symmetry groups acting on the coupling.

\subsection*{Symmetric Pad\'e approximants}

Consider a general situation in which a theory has a weak coupling expansion in some variable $\x$.\footnote{In \S\ref{s3} we will identify $\x$ with essentially the square of the Yang-Mills coupling constant, see Eqn.\,\eqref{gyparams}. The discussion here is more general, and $\x$ can represent an arbitrary parameter.} 
The theory may contain several other parameters, \eg, the theta angle, but we assume that the coefficients of the Taylor series do not depend on them. The $[n/m]$ Pad\'e approximant to such a function is the rational function
\be\label{Pade}
P_{[n/m]}(\x)=\frac{a_0+a_1\,\x+a_2\,\x^2+\cdots+a_n\,\x^n}{b_0+b_1\,\x+b_2\,\x^2+\cdots+b_m\,\x^m}~,
\ee
where the coefficients $\{a_k, b_k\}$ are chosen so that the Taylor series around $\x=0$ matches the known perturbative expansion to order $\x^{m+n+1}$. In general, one has an assortment of choices for the integers $n$ and $m$ that all allow for matching the same number of coefficients in the Taylor series.

We are concerned with the situation in which the function in question obeys a symmetry of the form $f(\x)=f({\bf h}\cdot \x)$ where ${\bf h}$ is a transformation of order $d$. As we will see below, ${\bf h}\cdot \x$ will generally depend not only on $\x$, but on the other parameters of the theory as well. Those other parameters will also transform under ${\bf h}$ into functions of themselves and $\x$. It is straightforward to build such a symmetry into the Pad\'e approximant by summing each monomial over images. Introducing the convenient notation
\be\label{sumoverimg}
\x_{\bf h}^n = \sum_{k=0}^{d-1}({\bf h}^k\cdot \x)^n~,
\ee
the symmetric Pad\'e approximant can be defined as\footnote{Despite the degenerate notation, the coefficients in Eqn.\,\eqref{Pade} and Eqn.\,\eqref{SPade} will not be the same for a given function.} 
\be\label{SPade}
P^{\,\bf h}_{[n/m]}(\x)=\frac{a_0\,\x_{\bf h}^{-n}+a_1\,\x_{\bf h}^{-n+1}+\cdots +a_{n-1}\,\x_{\bf h}^{-1}+ d\, a_n}
{b_0\,\x_{\bf h}^{-m}+b_1\,\x_{\bf h}^{-m+1}+\cdots+b_{m-1}\,\x_{\bf h}^{-1}+ d\, b_m}~,
\ee
where $\{a_k, b_k\}$ are again determined by requiring that the Taylor series expansion of \eqref{Pade} in $\x$ matches the known weak coupling expansion. For the symmetries used in this paper, ${\bf h}^{\ell}\cdot \x$ will always diverge as $C/\x$ for some constant $C$ as $\x\to 0$, so $P^{\,\bf h}_{[n/m]}\simeq a_0 \x^{m-n}/b_0$ for small $\x$. In this scenario, $n$ and $m$ must be chosen to reproduce the correct weak coupling behavior; our expansion will always begin at order $\x$, which then requires that we take $n=m-1$. After accounting for the freedom to simultaneously rescale all coefficients, there will be $m+n+1=2m$ independent coefficients $\{a_k, b_k\}$ that should be fixed by demanding that \eqref{Pade} correctly reproduce the known perturbative expansion up to and including terms of order $\x^{2m}$.

There is a subtlety related to the prescription outlined above. Because ${\bf h}^k\cdot \x$ may depend on additional parameters, the coefficients $\{a_k, b_k\}$ determined using this procedure can acquire a parameter-dependence. Since these parameters transform non-trivially under duality, Eqn.\,\eqref{SPade} will no longer necessarily be duality invariant. If the dependence of ${\bf h}^k\cdot \x$ on these additional parameters arises at order $\x^\ell$, then the coefficients of the Taylor series expansion of \eqref{SPade} will depend on them starting at order $\x^{m+\ell}$. For all the cases investigated in this paper we have one additional parameter, namely the theta angle, and $({\bf h}\cdot \x)^{-1}$ will depend on $\theta$ starting at order $\x^3$. Thus the first coefficient of the expansion of \eqref{SPade} that will depend on $\theta$ will arise at order $\x^{m+3}$.  This is a higher order than $\x^{2m}$ for $m\le 2$, so if we restrict ourselves to $m\leq2$ (corresponding to matching perturbation theory to at most order $\x^4$) then the coefficients $\{a_k, b_k\}$ will be independent of $\theta$ and we shall be free of this issue. Since at present the anomalous dimensions we study are only available to four loops, our analysis will be unaffected.\footnote{The curious reader may note that the requirement of perturbative $\theta$-independence is precisely the reason for the somewhat nonstandard negative exponents in the interpolations \eqref{SPade}--\eqref{etildefraction}.}

If instead of working in the full parameter space we choose to work on a one dimensional subspace that is invariant under ${\bf h}$, {\it e.g.} the imaginary axis in the upper half plane parameterized by $\tau$ for the choice ${\bf h}\cdot\tau=-1/\tau$, then on this line ${\bf h}$ takes $\x$ to a function of $\x$ only. In this case the difficulties mentioned above are absent and we can apply this procedure without concern. The price we pay is that the resulting interpolation will only be a plausible approximation of the desired function on this subspace.
 
We can also define an interpolating function that is related to an odd-degree Pad\'e approximant,
\be\label{SPadeHalf}
\widetilde{P}^{\,\bf h}_{[n/m]}(\x)=\frac{\tilde a_0\,\x_{\bf h}^{-n-\frac{1}{2}}+\,\cdots\, +
\tilde a_{n}\,\x_{\bf h}^{-\frac 12}}{\tilde b_0\,\x_{\bf h}^{-m-\frac{1}{2}}+\,\cdots\, +\tilde b_{m}\,\x_{\bf h}^{-\frac 12}}~.
\ee
Again, the cases of interest will require $n=m-1$, leaving $2m$ independent parameters that are fixed by matching Taylor series to order $\x^{2m}$. It can easily be seen that if the symmetry in question is order two and acts as ${\bf h}\cdot \x=k/\x$ for constant $k$, then $P^{\,\bf h}_{[n/m]}(\x)$ and $\widetilde P^{\,\bf h}_{[n/m]}(\x)$ are identical functions. Nevertheless we have introduced them separately here because in the general case, they will be inequivalent.

\subsection*{Fractional power of a polynomial}

Another resummation procedure that can be tailored for compatibility with S-duality was introduced in \cite{1304.0458}. For the sort of duality invariant function discussed above, we define the interpolation
\be \label{efraction}
F_n(\x)=\left(f_1\,\x_{\bf h}^{-\frac{2n-1}{2}}+f_2\,\x_{\bf h}^{-\frac{2n-3}{2}}
+\,\cdots\,+f_n\,\x_{\bf h}^{-\frac12}\right)^{-\frac{2}{2n-1}}~,
\ee
where the coefficients $\{f_k\}$ are again fixed by matching Taylor series around $\x=0$ to order $\x^n$. We refer to this as the fractional power of polynomial (FPP) approximation. We can also define an analogous integral-power version of this interpolation,
\be \label{etildefraction}
\widetilde{F}_n(\x)=\left(\tilde f_1\,\x_{\bf h}^{-(n-1)} + \tilde f_2\,\x_{\bf h}^{-(n-2)}+\,\cdots\,
+ \tilde f_{n-1}\,\x_{\bf h}^{-1}+ \tilde f_n\right)^{-\frac{1}{n-1}}~.
\ee
In all the cases we consider, $|{\bf h}^\ell\cdot \x| > C/\x$ as $\x\to 0$ with $C$ a positive constant. Consequently, the Taylor series expansion coefficients of $F_n(\x)$ or $\widetilde F_n(\x)$ up to order $\x^n$ are unchanged if we replace $\x_{\bf h}^{-k}$ by $\x^{-k}$ in the original expression. As a result, the coefficients $\{f_k\}$ and $\{\tilde f_k\}$, determined by matching the Taylor series expansion to order $\x^n$, are independent of the choice of ${\bf h}$, and hence of any other parameters in the theory. For this reason, in contrast to the symmetric Pad\'e approximant, there is no obstruction to using these interpolating functions to arbitrarily high order approximation. Another advantage enjoyed by FPP over the symmetric Pad\'e approximants is that FPP can be used to match a perturbative series to any loop order, even or odd, by appropriate choice of $n$. The Pad\'e approximants are limited to matching results at even loops orders. Unlike Pad\'e approximant, the two versions of FPP differ even for the case ${\bf h}\cdot \x=k/\x$.

\section{Application to anomalous dimensions in $\NN=4$ SYM}\label{s3}

The anomalous dimensions of local operators in $\NN=4$ SYM are real functions of the complex coupling constant
\be 
\tau = {\theta\over 2\pi} + {4\pi i\over \gym^2}~,
\ee
where $\gym$ is the Yang-Mills coupling and $\theta$ is the theta angle. Under an S-duality transformation corresponding to an element ${\bf h}\in {\rm PSL}(2,\mathbb{Z})$, the coupling transforms as
\be\label{edual}
{\bf h}\cdot\tau=\frac{a\tau+b}{c\tau+d}~,
\ee
with $a,b,c,d\,\in\,\mathbb{Z}$ satisfying $ad-bc=1$. For later convenience of notation, we define
\be\label{gyparams}
g := {\gym^2 \over 4\pi}~, \qquad y := {\theta\over 2\pi}~,
\ee
and denote the corresponding transformations of $g$ and $y$ as ${\bf h}\cdot g$ and ${\bf h}\cdot y$. The anomalous dimensions have perturbative expansions of the form
\be
\gamma_{\rm pert}(g) = \sum_{n=1}^\infty \gamma_ng^n~,
\ee
where the coefficient $\gamma_n$ can be computed, \eg, from Feynman diagrams with $n$ loops. Although there is no $y$ dependence at any order in perturbation theory, the non-perturbative functions $\gamma(g,y)$ will in general depend on $y$. In our interpolations, $y$ dependence will be introduced automatically by the requirement of duality invariance.

We apply the prescriptions of the previous section to define interpolating functions that match $\gamma_{\rm pert}(g)$ to a given order in perturbation theory around a weak-coupling limit of $\tau$, and that are invariant under finite-order subgroups of PSL$(2,\BBZ)$. Up to conjugation, there are two such subgroups, each of which fixes a single point on the upper half plane. Without loss of generality, we can restrict our attention to the canonical fundamental domain of the modular group, within which these fixed points occur at
\be
\tau_{2} = i~,\qquad\qquad \tau_{3}=\exp (i\pi/3)~.
\ee
The point $\tau_{2}$ is invariant under the order-two electric/magnetic duality transformation,
\be\label{EMtrans}
{\bf S}\cdot\tau = -\frac{1}{\tau}~,
\ee
while the point $\tau_{3}$ is invariant under the order three transformation
\be\label{Z3trans}
({\bf T\cdot S})\cdot\tau = {\tau-1\over\tau}~.
\ee
Some consideration is necessary to decide in what regions of the upper half plane the corresponding interpolations have the potential to be good approximations to the anomalous dimensions. The true anomalous dimensions will be modular functions, and so will obey many relations on the upper half plane. Because our approximations only take into account a finite number of these relations, we have no right to expect any accuracy in a generic region of the upper half plane. They should, however, be best suited for approximating the values of anomalous dimensions at the corresponding fixed point, as well as within the copies of the fundamental domain to which the fixed point belongs (see Figs.~\ref{fig:Z2fig} and \ref{fig:Z3fig}).

\ZtwoUHP
\subsection*{$\mathbb{Z}_2$ invariant interpolation}

The basic S-duality operation of Eqn.\,\eqref{EMtrans} acts on the upper half plane as a reflection through the unit semi-circle along with a reflection through the imaginary axis $y\leftrightarrow -y$. The induced actions on the Yang-Mills coupling and theta angle are
\be\label{edual1}
{\bf S}\cdot g ={1+y^2g^2\over g} ~,\qquad {\bf S}\cdot y=-{y\, g^2\over1+y^2g^2}~.
\ee
In particular, this transformation sends the line at $\theta=0$ to itself via a reflection through the fixed point $\tau=\tau_2$.

For the order two subgroup generated by $\bf S$, the sum over images
\eqref{sumoverimg} becomes 
\be\label{S1rep}
\x_{\bf S}^k = g^k + \left({1+y^2g^2\over g}\right)^k~.
\ee
The resulting interpolating functions are manifestly invariant under \eqref{EMtrans}, and after fixing the coefficients appropriately they will have the correct perturbative expansions about $\tau=0$ and $\tau=0+i\infty$. In addition, from Eqn.\,\eqref{S1rep} it is clear that the resulting function will be invariant under $y\leftrightarrow -y$, which is required by CP invariance of the operators involved.

In the most optimistic scenario, these interpolating functions may give a good approximation to the anomalous dimensions in the shaded region of Fig.~\ref{fig:Z2fig}, with the best case likely being the dark line at $\theta=0$. We will consequently use these interpolations primarily to study the fixed point at $\tau=\tau_2$, with the other fixed point at $\tau=\tau_3$ being a borderline case.

\ZthreeUHP

\subsection*{$\mathbb{Z}_3$ invariant interpolation}

Up to conjugation, the unique $\BBZ_3$ subgroup of PSL$(2,\BBZ)$ is generated by the transformation \eqref{Z3trans}, which acts on $(g,y)$ as 
\be
({\bf T}\cdot{\bf S})\cdot g = {1 + y^2g^2 \over g}, \qquad  ({\bf T}\cdot{\bf S})\cdot y = 1 -
{g^2 y\over 1+y^2g^2}~.
\ee 
This symmetry permutes the three dark segments in Fig.~\ref{fig:Z3fig}, fixing the junction where they intersect at $\tau=\tau_3$.

In this case the interpolating functions are obtained from the corresponding building block:
\be
g_{\bf T \cdot \bf S}^k =  g^k+\left({1+y^2g^2\over g}\right)^k+\left({1+(1-y)^2g^2)\over g}\right)^k~.
\ee
After fixing the coefficients appropriately, the interpolating functions so-defined are guaranteed to have the correct perturbative expansions around $\tau=0$, $\tau=1$, and $\tau=\frac12+i\infty$. In addition, they have the correct symmetry structure at the fixed point $\tau=\tau_3$, along with the correct invariance under $y\leftrightarrow 1-y$. Such an interpolating function has the chance to yield a good approximation to the true, modular invariant function in the shaded regions of Fig.~\ref{fig:Z3fig}, with the most compelling loci being the dark purple lines. We will use this resummation to estimate the values of anomalous dimensions primarily at $\tau=\tau_3$, with the value at $\tau_2$ also being of interest.

\spinzeroSUtwo
\spinzeroSUthree
\spinzeroSUfour
\spintwoSUtwo
\spintwoSUthree
\spintwoSUfour

\section{Results}\label{s4}
We use the interpolations described above to approximate the anomalous dimensions of operators of the form
\be\label{singletraceops}
\OO_M=\Tr\,\phi^ID^M \phi^I~,\qquad M=0,2,4,\cdots~.
\ee
in $\NN=4$ SYM with gauge group $SU(N)$. These operators are $SU(4)_R$ singlets and superconformal primary operators belonging to long representations of the superconformal algebra. They have perturbative scaling dimensions $\Delta_M = 2 + M + \gamma_M(g)$, where the perturbative anomalous dimension $\gamma_M(g)$ is independent of the theta angle. The anomalous dimensions have been computed by a variety of methods to quite high orders in perturbation theory. In what follows we will restrict ourselves to the gauge groups $SU(2)$, $SU(3)$ and $SU(4)$, where we expect the best performance from our interpolation methods (\cf\ \S\ref{s5}).

\subsection*{The Konishi operator}\label{s4.1}

We begin with the result for $M=0$, which corresponds to the Konishi operator. In an impressive series of papers \cite{TwistTwo,5LoopKonishi}, the Konishi anomalous dimension has been computed up to four loop order,
\be
\gamma_0(g) =
\tfrac{3Ng}{\pi}-\tfrac{3N^2g^2}{\pi^2}+\tfrac{21N^3g^3}{4\pi^3}+\left(-39 + 9\,\zeta(3) - 45\,\zeta(5)\left(\tfrac{1}{2}+\tfrac{6}{N^2}\right)\right)\tfrac{N^4g^4}{4\pi^4}+\cdots~,
\ee
where we recall that $g=\gym^2/4\pi$.

We have applied the interpolation techniques of \S\ref{s3} to estimate the function $\gamma_0(g,y)$ in various regions of the upper half plane. In Figs.~\ref{fig:Spin0SU2}--\ref{fig:Spin0SU4}, we present the resulting functions evaluated along the interesting one-dimensional subspaces of the upper half-plane. Of particular interest are the values at the fixed points $\tau_2$ and $\tau_3$, which are stationary points of the anomalous dimensions, and so are the most likely candidates for saturating the bounds of \cite{1304.1803}.

\subsection*{Spin two operator}\label{s4.2}

Next we consider the case of the $M=2$ operator, whose perturbative anomalous dimension is given by \cite{TwistTwo,NPSpin2}
\be
\gamma_2(g)=\tfrac{25Ng}{6\pi}-\tfrac{925N^2g^2}{216\pi^2}+\tfrac{241325N^3g^3}{31104\pi^3}+(\gamma_2^{\textsc{ABA}}+\gamma_2^{wrap}+\gamma_2^{np})\tfrac{g^4N^4}{(4\pi)^4}+\cdots~,
\ee
with
\bea
&\gamma_2^{\textsc{ABA}} &= -\tfrac{304220675}{69984}-\tfrac{3250\,\zeta(3)}{9} ~,\\
&\gamma_2^{wrap} &= \tfrac{5196875}{7776}+\tfrac{143750\,\zeta (3)}{81}-\tfrac{25000 \,\zeta(5)}{9}~, \\
&\gamma_2^{np} &= \tfrac{8400+28000\,\zeta(3)-100000\,\zeta(5)}{3N^2}~.
\eea
We can repeat the analysis of the previous subsection for this case; the results are shown in Figs.~\ref{fig:Spin2SU2}--\ref{fig:Spin2SU4}.

\subsection*{Spin four operator}\label{s4.3}

The anomalous dimension of the $M=4$ operator is given by \cite{TwistTwo}
\be
\gamma(g) = \tfrac{49Ng}{10\pi}-\tfrac{45619N^2g^2}{9000\pi^2}+\tfrac{300642097N^3g^3}{32400000\pi^3}+\left(\gamma_4^{\textsc{ABA}}+\gamma_4^{wrap}+\gamma_4^{np}\right)\tfrac{g^4N^4}{(4\pi)^4}+\cdots~,
\ee
where
\be
\gamma_4^{\textsc{ABA}} = -\tfrac{1916919629681}{364500000}-\tfrac{91238\,\zeta(3)}{225}~,\quad\gamma_4^{wrap} = \tfrac{2526915643}{2700000}+\tfrac{4672346\,\zeta (3)}{1875}-\tfrac{19208\,\zeta(5)}{5}~.
\ee
To the best of our knowledge, the non-planar contribution $\gamma_4^{np}$ has not yet been calculated. As a result, we cannot find the interpolating functions to four loops. The results up to three loops are similar to those for spin zero and spin two operators, but we do not display them here. Numerical results for the values of the spin four interpolating function at duality fixed points are presented in Table \ref{t2}.

\subsection*{Comments on interpolations}\label{s4.4}

A few immediate comments are in order regarding the behavior of the interpolating functions. 

Upon examination of Figs.~\ref{fig:Spin0SU2}--\ref{fig:Spin2SU4}, we see that for a given choice of duality subgroup, loop order, and region of evaluation, there is very good agreement between the different interpolating functions (two Pad\'e and two FPP). For example, for the Konishi interpolation with $SU(2)$ gauge group, two prescriptions never differ from their mean by more than 2.5\% over the full range of values of $g$ for $\theta=0$ or $\theta=\pi$. This is encouraging, because the interpolating functions have been chosen using the somewhat capricious criterion of ``simplicity'', rather than a specific physical motivation. It is a positive sign that the results do not depend heavily upon exactly what function is used, at least within the small family of functions we have tested.

On the other hand, there is a decent amount of variation between the different loop orders -- especially between the two loop and three loop results. This is not unexpected, but the size of the variation makes it clear that one should expect the next correction to still be nontrivial. An optimist may note that the four loop result lies between the two and three loop results, and this may be the start of an alternating progression that converges towards the actual anomalous dimension.

Finally, there is a distinction to be drawn between the cases in which the interpolating function is evaluated at the boundary of its domain of conjectured validity (plots appearing in the bottom-left and top-right corners of the respective figures) and the cases for which the function is evaluated along its optimal locus (top-left and bottom-right corners). In particular, the evaluation of a $\BBZ_2$ invariant resummation at $\tau=\tau_3$ is likely to yield an overestimate: the actual anomalous dimension at $\tau_3$ should be at a stationary point, while in the interpolation it is only stationary along the radial direction, and is still increasing in the $g$ and $y$ directions. The situation is somewhat better in the case of extrapolating the $\BBZ_3$ invariant resummations to $\tau_2$. In this case, the combination of $\BBZ_3$ symmetry with invariance under $y\leftrightarrow1-y$ guarantees that at $y=0$, the interpolating function is invariant under $g\leftrightarrow 1/g$, making the result stationary along the $g$ axis at $\tau_2$. However, the interpolation does not account for $y\leftrightarrow-y$ invariance, which enforces stationarity in the $y$ direction at $\tau_2$. Indeed, the superior accuracy on the optimal locus can be checked explicitly by applying these interpolations to known modular invariant functions, such as the real Eisenstein series.

\spinZeroTwoTable
\spinFourTable

For the purpose of comparing our resummations to the results of \cite{1304.1803}, we focus on the values taken by the interpolated anomalous dimensions at the fixed points $\tau=\tau_{2,3}$. The results for spin zero and spin two are summarized in Table \ref{t1}. In order to assign a single value -- with error bar -- to a fixed point requires some artistry. We have chosen to draw the values for $\tau_2$ from the $\BBZ_2$-invariant resummations, and the $\tau_3$ values from the $\BBZ_3$-invariant resummations. In particular, we take the mean of the two loop results as the lower end of the error bar, and the mean of the three loop results as the upper end. For a central value, we take a weighted average of the mean values for each loop order. The weights are given by the inverse of the spread in values at that order. With this choice of the error bar, we find that at $\tau=\tau_2$, all the four loop results from $\BBZ_2$ and $\BBZ_3$ invariant interpolating functions lie within the error bars. At $\tau=\tau_3$ all the four loop results from $\BBZ_3$ invariant interpolating functions lie within the error bar. The four loop $\BBZ_2$ invariant interpolating functions at $\tau_3$ 
lie near the top of the range and occasionally overshoots the upper limit, but even the maximum violation is quite small ($\sim .05$). Table \ref{t2} gives the corresponding results for spin four operators. However, in the absence of the four loop results, we can only give the range in which the anomalous dimension is expected to lie, the lower and upper limits being the average two and three loop results respectively.

\figureconformal

For every point $\tau$ on the conformal manifold, there is a set of numbers $(\Delta_0, \Delta_2, \Delta_4, \cdots)$ describing the dimensions of the lowest twist operators of spin zero, two, and four. We expect that under this map, the fundamental domain of the conformal manifold will trace out a two dimensional subspace in $(\Delta_0, \Delta_2, \Delta_4, \cdots)$ space. Using our interpolation formula, we can try to identify the projection of this subspace to the $(\Delta_0, \Delta_2)$ plane. Due to $y\to -y$ symmetry, we can focus on the region bounded by the curves $y=0$, $y=1/2$ and $y^2 + g^{-2}=1$. To keep the analysis simple we use a weighted average of the interpolating functions, 
\be
{1\over 6}\, (\text{two loop average + 2 $\times$ three loop average + 3 $\times$ 
four loop average})~.
\ee
On the $y=0$ axis we use the $\BBZ_2$ invariant interpolation formula, on the $y=1/2$ axis we use the $\BBZ_3$ invariant interpolation function and on the circle $y^2 + g^{-2}=1$ we use a linear combination of these two which varies from being the $\BBZ_2$ invariant function at $\tau=i$ to $\BBZ_3$ invariant function at $\tau=\exp(i\pi/3)$. By tracing out the images of these boundaries in the $(\Delta_0, \Delta_2)$ plane we encounter a surprise: instead of forming the boundary of a two dimensional region they appear to lie along a one dimensional curve -- in fact a straight line with slope $25/18$ determined by the one loop anomalous dimension. This is shown in Fig.~\ref{fconformal}. Furthermore, this result seems to be quite robust; if we use the $\BBZ_2$ (or $\BBZ_3$) invariant interpolation for all the boundaries, we get essentially the same result with the same straight line; the only difference being that the line extends a little further (or less far) at the upper end. This result is also quite robust under the change in the averaging procedure; if we had used only the average two, three, or four loop results, we would get more or less the same curve except for a tiny deviation at the top. Thus this result seems to be much less uncertain compared to the actual values of the anomalous dimensions at $\tau_2$ and $\tau_3$. We have also checked that all points in the interior of the fundamental domain and not just on the boundary lie on the same straight line.

We are, of course, not suggesting the $(\Delta_2-4)/(\Delta_0-2) = 25 /18$ will hold as an exact relation -- this would be inconsistent with perturbation theory. Moreover the non-perturbative $\theta$-dependence will introduce a finite width.\footnote{While the two-point function of the Konishi operator is known to not receive instanton corrections \cite{Bianchi}, we still expect  the anomalous dimension of the correct eigenstate to be $\theta$-dependent in the full quantum theory.
}
However, we have checked that throughout the fundamental region the ratio stays very close to $25/18$ -- the maximum deviation being of the order of $0.6\%$. Thus what our result indicates is that the whole conformal manifold maps to a very narrow band in the $(\Delta_0, \Delta_2)$ plane. We believe similar results will continue to hold for other $\Delta_m$'s as well, indicating that the conformal manifold maps to a very narrow strip around a straight line in $\Delta_m$ space. Amusingly, for $SU(2)$ gauge group, the ratios of the corner values of the anomalous dimensions of spin zero and spin two operators (\cf\ Table \ref{t1}) is $1.28/.93$, which lies within 1\% of the ratio 25/18. This suggests that this may actually represent a physically realizable point. For $SU(3)$ and $SU(4)$, the agreement is not so good, which may indicate
that the interpolation method is not reliable for higher rank gauge groups.

\bootstrapCompare

\section{Discussion}\label{s5}

We have seen that one may obtain reasonable, self-consistent results by performing simple, duality-invariant resummations of perturbative anomalous dimensions in $\NN=4$ $SU(N)$ SYM. Probably the most interesting aspect of these interpolations is their relation to the results of the conformal bootstrap program for $\NN=4$ SYM \cite{1304.1803}. We recall that in that work, absolute bounds were derived for the anomalous dimensions of the first operator of spin zero, two, and four appearing in the OPE of a certain four point function. These are the bounds that are displayed under the heading `Bound' in Tables \ref{t1} and \ref{t2}. However, a more subtle result was obtained by tracing out the boundary between operator spectra that could be excluded by the conformal bootstrap and those that could not, parameterized by the values of those anomalous dimensions. In the three-dimensional octant spanned by the dimensions of the spin zero, two, and four operators, this boundary was found to be approximately cube-shaped, leading to the natural conjecture that the actual operator dimensions at a self-dual point can be obtained from the point at the corner. By estimating the location of the corner, which due to the numerical methods sits a little bit below the actual bounds, we obtain an improved estimate of the value of the anomalous dimensions at one of the S-duality fixed points on the $\NN=4$ SYM conformal manifold (there is no way to tell which one). Representative values of these estimates are displayed under the heading `Corner' in the tables.

In Fig.~\ref{fig:comparison}, we show the results of our interpolations relative to the boundary separating admissible spectra from inadmissible ones in the space of spin zero and spin two anomalous dimensions. For a given gauge group, spectra outside the approximately square regions are excluded. We see that for low $N$, the interpolations are in good agreement with the conjecture that the bounds are saturated at one of the duality fixed points. The quality of the agreement diminishes with increased $N$, but this comes as no surprise; as $N$ increases the effective coupling constant $gN$ takes larger values at the self dual point $g=1$, rendering perturbation theory and S-duality  
insufficient to control the behavior of the function everywhere. Moreover, for large enough $N$ we expect the anomalous dimensions of the studied operators to grow large for there to be substantial mixing with other operators, \eg, the Konishi operator will mix with a double-trace operator of tree level dimension four. This should lead to new features in the behavior of the anomalous dimensions as a function of the coupling that do not follow from naive extrapolation of the behavior at weak coupling. For these reasons, if there are general lessons to be learned that hold for all gauge groups, we are most likely to discover them by studying the results for the $SU(2)$ gauge group. 

\acknowledgments

We wish to thank
Nikolay Bobev,
Davide Gaiotto,
Rajesh Gopakumar,
Romuald Janik,
Dileep Jatkar,  
Gregory Korchemsky,
Peter Koroteev,
Hugh Osborn, 
Jo\~ao Penedones, 
Ricardo Schiappa,
Amit Sever, 
and Vitaly Velizhanin 
for useful discussions and correspondence. The work of B.v.R. and L.R. is partially supported by the NSF under Grants PHY-0969919 and PHY-0969739. The work of A.S. is supported in part by DAE project 12-R\&D-HRI-5.02-0303 and the J.C. Bose fellowship of DST, Govt. of India. C.B. would like to Perimeter Institute for Theoretical Physics for hospitality during the completion of this work. Research at Perimeter Institute is supported by the Government of Canada through Industry Canada and by the Province of Ontario through the Ministry of Economic Development \& Innovation.
\appendix
\section{Explicit interpolation formul\ae}\label{sa}

In this appendix, we provide the explicit interpolation formul\ae\ we have used in this 
paper. Suppose the perturbative expansion of the anomalous dimension of an operator
takes the form
\be
\gamma(g) = a\, g\,  ( 1 + b\, g^2 + c\, g^3 + d\, g^4 + \OO(g^5))\, .
\ee
Then the various interpolations are given below. 

\subsubsection*{$\BBZ_2$ invariant interpolation for FPP with half-integral powers}
\bea
\text{Two loops} &:& a\, \Bigg[ \bigg\{\frac{1}{g^{3/2}}-\frac{3 b}{2 \sqrt{g}}\bigg\} + \bigg\{ g\to {(1 + y^2 g^2) \over g}\bigg\}  \bigg]^{-2/3}
\nonumber \\
\text{Three loops} &:& a\, \Bigg[ \bigg\{
\frac{1}{g^{5/2}}-\frac{5 b}{2 g^{3/2}}+\frac{\frac{35 b^2}{8}-\frac{5
   c}{2}}{\sqrt{g}}\bigg\} + \bigg\{ g\to {(1 + y^2 g^2) \over g}\bigg\} \bigg]^{-2/5}
\nonumber \\
\text{Four loops} &:& a\, \Bigg[\Bigg\{ \frac{1}{g^{7/2}}-\frac{7 b}{2 g^{5/2}}+\frac{\frac{63 b^2}{8}-\frac{7
   c}{2}}{g^{3/2}}-\frac{7 \left(33 b^3-36 b c+8 d\right)}{16
   \sqrt{g}}\Bigg\}
      + \Bigg\{ g\to {(1 + y^2 g^2) \over g}\Bigg\} \Bigg]^{-2/7} \nonumber \\
\eea
\subsubsection*{$\BBZ_2$ invariant interpolation for FPP with integral powers}
\bea
\text{Two loops} &:& a\, \Bigg[ \bigg\{\frac{1}{g}\bigg\}  
+ \bigg\{ g\to {(1 + y^2 g^2) \over g}\bigg\} -b \bigg]^{-1}
\nonumber \\
\text{Three loops} &:& a\, \Bigg[ \bigg\{ \frac{1}{g^2}-\frac{2 b}{g}
\bigg\} + \bigg\{ g\to {(1 + y^2 g^2) \over g}\bigg\} +3 b^2 - 2 c\bigg]^{-1/2}
\nonumber \\
\text{Four loops} &:& a\, \Bigg[\Bigg\{\frac{1}{g^3}-\frac{3 b}{g^2}+\frac{6 b^2-3 c}{g}\Bigg\}
     + \Bigg\{ g\to {(1 + y^2 g^2) \over g}\Bigg\}  
+\left(-10 b^3+12 b c-3
   d\right)\Bigg]^{-1/3} \, . \nonumber \\
\eea

\subsubsection*{$\BBZ_2$ invariant interpolation for Pad\'e with half-integral powers}
\bea
\text{Two loops} &:&  { a} \Bigg[ \bigg\{{1\over g^{1/2}}\bigg\}  + \bigg\{ g\to {(1 + y^2 g^2) \over g}\bigg\}  \bigg\}\bigg] \nonumber \\ &&
\Bigg[
\left\{\frac{1}{g^{3/2}}-\frac{b-1}{\sqrt{g}}\right\} + \bigg\{ g\to {(1 + y^2 g^2) \over g}\bigg\} \Bigg]^{-1}
 \nonumber \\
\text{Four loops} &:& a \Bigg[ \left\{\frac{b^3+ b^2-2 b c- c+d-1}{\sqrt{g}
   \left(b^2-c-1\right)}+\frac{1}{g^{3/2}}\right\}
+ \bigg\{ g\to {(1 + y^2 g^2) \over g}\bigg\} 
\Bigg] \nonumber \\ &&
\Bigg[
\bigg\{\frac{1}{g^{5/2}} + \frac{b^2-b
   c+b-c+d-1}{g^{3/2} \left(b^2-c-1\right)}+\frac{-b c-b d+b+c^2+d-1}{\sqrt{g} \left(b^2-c-1\right)}\bigg\} \nonumber \\
&& 
+ \bigg\{ g\to {(1 + y^2 g^2) \over g}\bigg\} 
\Bigg]^{-1} 
\eea

\subsubsection*{$\BBZ_2$ invariant interpolation for Pad\'e with integral powers}
\bea
\text{Two loops} &:&  { a} \Bigg[ \bigg\{{1\over g}\bigg\}  + \bigg\{ g\to {(1 + y^2 g^2) \over g}\bigg\} - b\bigg]^{-1} \nonumber \\
\text{Four loops} &:& a \Bigg[ \bigg\{\frac{1}{g}\bigg\} 
+ \bigg\{ g\to {(1 + y^2 g^2) \over g}\bigg\} 
+ \frac{b^3-2 b c+d}{b^2-c-1}\Bigg] \nonumber \\ &&
\Bigg[
\bigg\{\frac{1}{g^2} + \frac{b-bc+d}{g
   \left(b^2-c-1\right)}\bigg\}
+ \bigg\{ g\to {(1 + y^2 g^2) \over g}\bigg\} 
 + \frac{c^2-b d-1}{b^2-c-1}
\Bigg]^{-1} 
\eea

\subsubsection*{$\BBZ_3$ invariant interpolation for FPP with half-integral powers}
\bea
\text{Two loops} &:& a\, \Bigg[ \bigg\{\frac{1}{g^{3/2}}-\frac{3 b}{2 \sqrt{g}}\bigg\} + \bigg\{ g\to {(1 + y^2 g^2) \over g}\bigg\} + \bigg\{ g\to {(1 + (1-y)^2 g^2) \over g}\bigg\} \bigg]^{-2/3}
\nonumber \\
\text{Three loops} &:& a\, \Bigg[ \bigg\{
\frac{1}{g^{5/2}}-\frac{5 b}{2 g^{3/2}}+\frac{\frac{35 b^2}{8}-\frac{5
   c}{2}}{\sqrt{g}}\bigg\} + \bigg\{ g\to {(1 + y^2 g^2) \over g}
   + \bigg\{ g\to {(1 + (1-y)^2 g^2) \over g}\bigg\} \bigg]^{-2/5}
\nonumber \\
\text{Four loops} &:& a\, \Bigg[\Bigg\{ \frac{1}{g^{7/2}}-\frac{7 b}{2 g^{5/2}}+\frac{\frac{63 b^2}{8}-\frac{7
   c}{2}}{g^{3/2}}-\frac{7 \left(33 b^3-36 b c+8 d\right)}{16
   \sqrt{g}}\Bigg\}  \nonumber \\
&&
      + \Bigg\{ g\to {(1 + y^2 g^2) \over g}\Bigg\} 
      + \Bigg\{ g\to {(1 + (1-y)^2 g^2) \over g}\Bigg\} \Bigg]^{-2/7} \, .
\eea
\subsubsection*{$\BBZ_3$ invariant interpolation for FPP with integral powers}
\bea
\text{Two loops} &:& a\, \Bigg[ \bigg\{\frac{1}{g}\bigg\}  
+ \bigg\{ g\to {(1 + y^2 g^2) \over g}\bigg\} + \bigg\{ g\to {(1 + (1-y)^2 g^2) \over g}\bigg\}
-b \bigg]^{-1}
\nonumber \\
\text{Three loops} &:& a\, \Bigg[ \bigg\{ \frac{1}{g^2}-\frac{2 b}{g}
\bigg\} + \bigg\{ g\to {(1 + y^2 g^2) \over g}\bigg\} + \bigg\{ g\to {(1 + (1-y)^2 g^2) \over g}\bigg\}
+3 b^2 - 2 c\bigg]^{-1/2}
\nonumber \\
\text{Four loops} &:& a\, \Bigg[\Bigg\{\frac{1}{g^3}-\frac{3 b}{g^2}+\frac{6 b^2-3 c}{g}\Bigg\}
     + \Bigg\{ g\to {(1 + y^2 g^2) \over g}\Bigg\}  + \Bigg\{ g\to {(1 + (1-y)^2 g^2) \over g}\Bigg\}
 \nonumber \\
&&+\left(-10 b^3+12 b c-3
   d\right)\Bigg]^{-1/3}
\eea

\subsubsection*{$\BBZ_3$ invariant interpolation for Pad\'e with half-integral powers}
\bea
\text{Two loops} &:&  { a} \Bigg[ \bigg\{{1\over g^{1/2}}\bigg\}  + \bigg\{ g\to {(1 + y^2 g^2) \over g}\bigg\} + \bigg\{ g\to {(1 + (1-y)^2 g^2) \over g} \bigg\}\bigg] \nonumber \\ &&
\Bigg[
\left\{\frac{1}{g^{3/2}}-\frac{b-2}{\sqrt{g}}\right\} + \bigg\{ g\to {(1 + y^2 g^2) \over g}\bigg\} + \bigg\{ g\to {(1 + (1-y)^2 g^2) \over g} \bigg\}
\Bigg]^{-1}
 \nonumber \\
\text{Four loops} &:& a \Bigg[ \left\{\frac{b^3+2 b^2-2 b c-2 c+d-2}{\sqrt{g}
   \left(b^2-c-4\right)}+\frac{1}{g^{3/2}}\right\}
+ \bigg\{ g\to {(1 + y^2 g^2) \over g}\bigg\} \nonumber \\ &&
+ \bigg\{ g\to {(1 + (1-y)^2 g^2) \over g}\bigg\}
\Bigg] \nonumber \\ &&
\Bigg[
\bigg\{\frac{1}{g^{5/2}} + \frac{2 b^2-b
   c+4 b-2 c+d-2}{g^{3/2} \left(b^2-c-4\right)}+\frac{-2 b c-b d+2 b+c^2+2 d-4}{\sqrt{g} \left(b^2-c-4\right)}\bigg\} \nonumber \\
&& 
+ \bigg\{ g\to {(1 + y^2 g^2) \over g}\bigg\} + \bigg\{ g\to {(1 + (1-y)^2 g^2) \over g}\bigg\}
\Bigg]^{-1} 
\eea

\subsubsection*{$\BBZ_3$ invariant interpolation for Pad\'e with integral powers}
\bea
\text{Two loops} &:&  { a} \Bigg[ \bigg\{{1\over g}\bigg\}  + \bigg\{ g\to {(1 + y^2 g^2) \over g}\bigg\} + \bigg\{ g\to {(1 + (1-y)^2 g^2) \over g} \bigg\}- b\bigg]^{-1} \nonumber \\
\text{Four loops} &:& a \Bigg[ \bigg\{\frac{1}{g}\bigg\} 
+ \bigg\{ g\to {(1 + y^2 g^2) \over g}\bigg\} + \bigg\{ g\to {(1 + (1-y)^2 g^2) \over g}\bigg\}
+ \frac{b^3-2 b c+d}{b^2-c-2}\Bigg] \nonumber \\ &&
\Bigg[
\bigg\{\frac{1}{g^2} + \frac{2 b-bc+d}{g
   \left(b^2-c-2\right)}\bigg\}
+ \bigg\{ g\to {(1 + y^2 g^2) \over g}\bigg\} + \bigg\{ g\to {(1 + (1-y)^2 g^2) \over g}\bigg\}
\nonumber \\
&& + \frac{c^2-b d-4}{b^2-c-2}
\Bigg]^{-1} 
\eea

Note that at the two loop order FPP with integral powers and Pad\'e approximant with
integral powers coincide both for $\BBZ_2$ and $\BBZ_3$ invariant interpolation. 
With a little bit of work one can also verify that on the $y=0$ line $\BBZ_2$ invariant
Pad\'e with integral and half-integral powers coincide both for two and four loops.
The latter coincidence has already been discussed in the main text.

\end{document}

%% file: resummation_figures.tex
\def\ZtwoUHP{
\begin{figure}%[t!]
\begin{center}
\epsfysize=7cm
\epsfbox{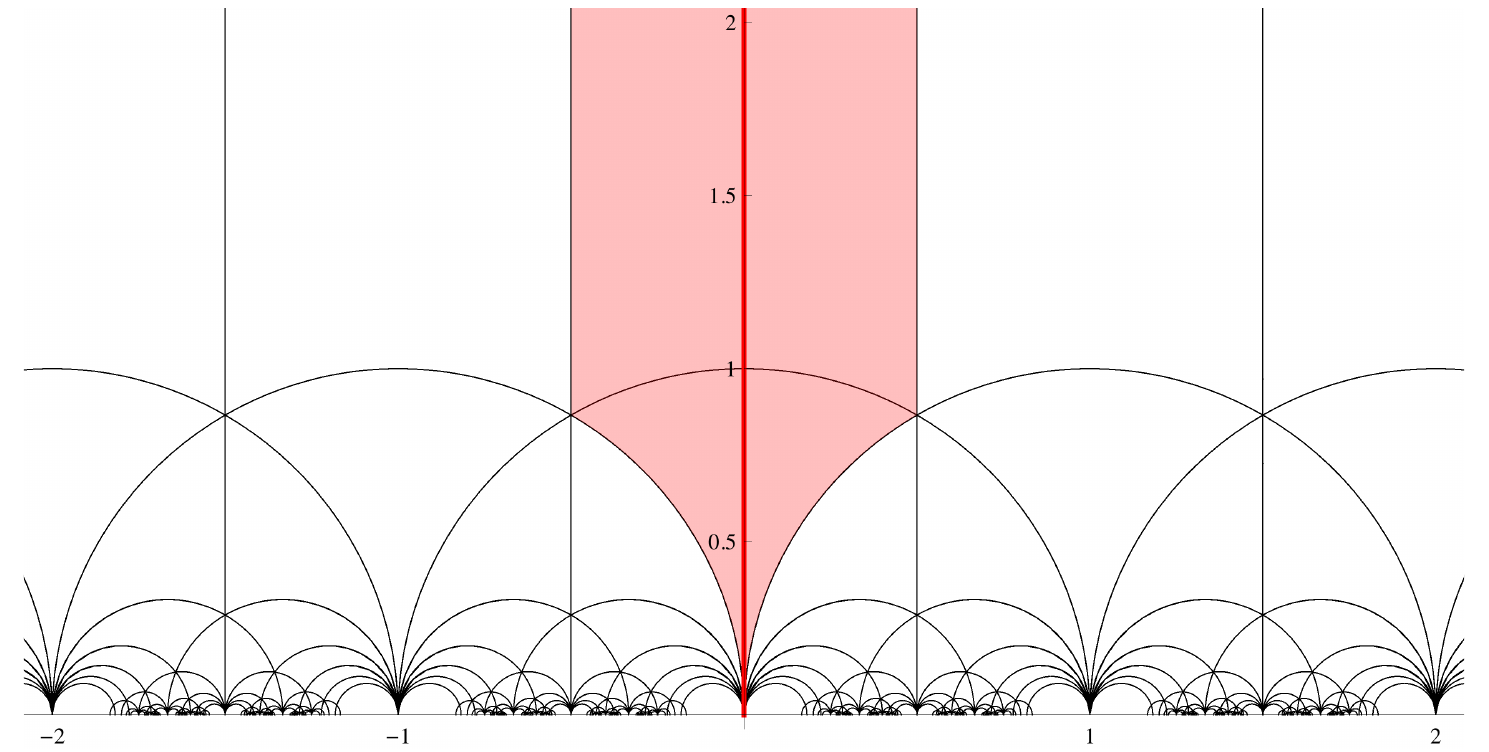}
\end{center}
\vspace{-20pt}
\caption{The upper half plane is tessellated by images of the fundamental domain of PSL$(2,\BBZ)$. The $\BBZ_2$-invariant interpolating functions defined here are well-suited to describe anomalous dimensions in two copies of the fundamental domain, shown as shaded in the figure. The solid line is $\theta=0$, and represents the best case for the 
$\BBZ_2$-invariant  interpolating function.
\label{fig:Z2fig}}
\end{figure}}

\def\ZthreeUHP{
\begin{figure}[t!]
\begin{center}
\epsfysize=7cm
\epsfbox{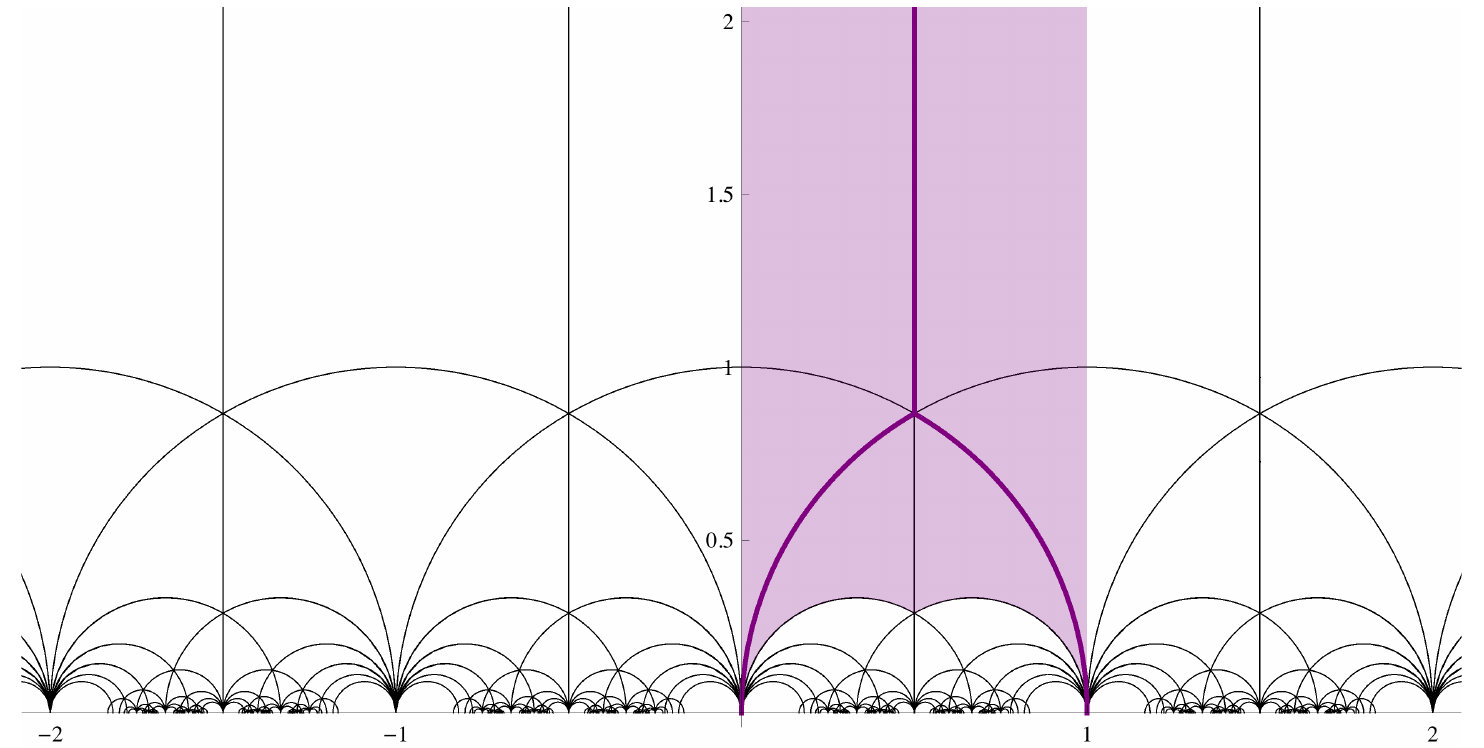}
\end{center}
\vspace{-20pt}
\caption{The $\BBZ_3$ invariant interpolation is particularly well suited to describe anomalous dimensions along the bold segments in the above figure. Because the method is not invariant under $y\leftrightarrow -y$, its accuracy is sure to degenerate for, \eg, $y<0$. The regions where the best behavior is expected are again shaded.\label{fig:Z3fig}}
\end{figure}}

\def\spinzeroSUtwo{
\begin{figure}[ht!]
\begin{center}
\hbox{
\epsfysize=4.85cm\epsfbox{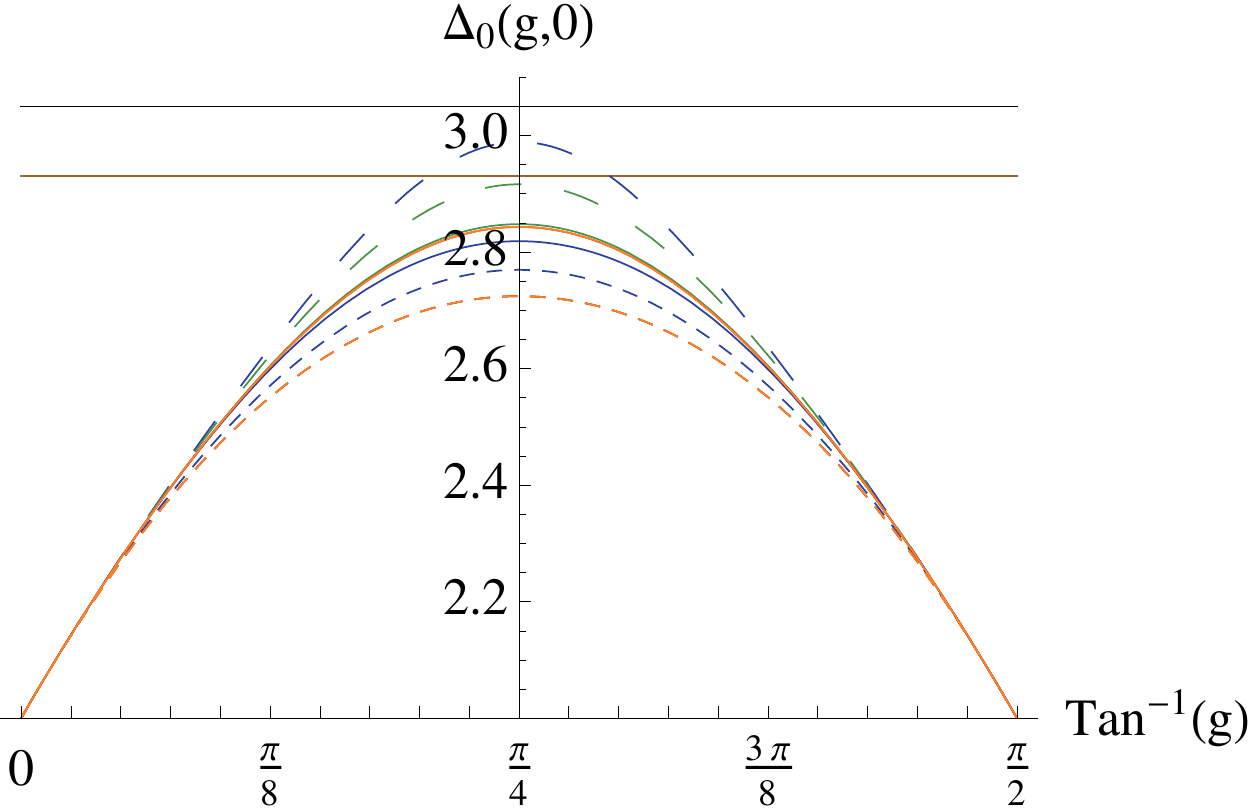}~~
\epsfysize=4.85cm\epsfbox{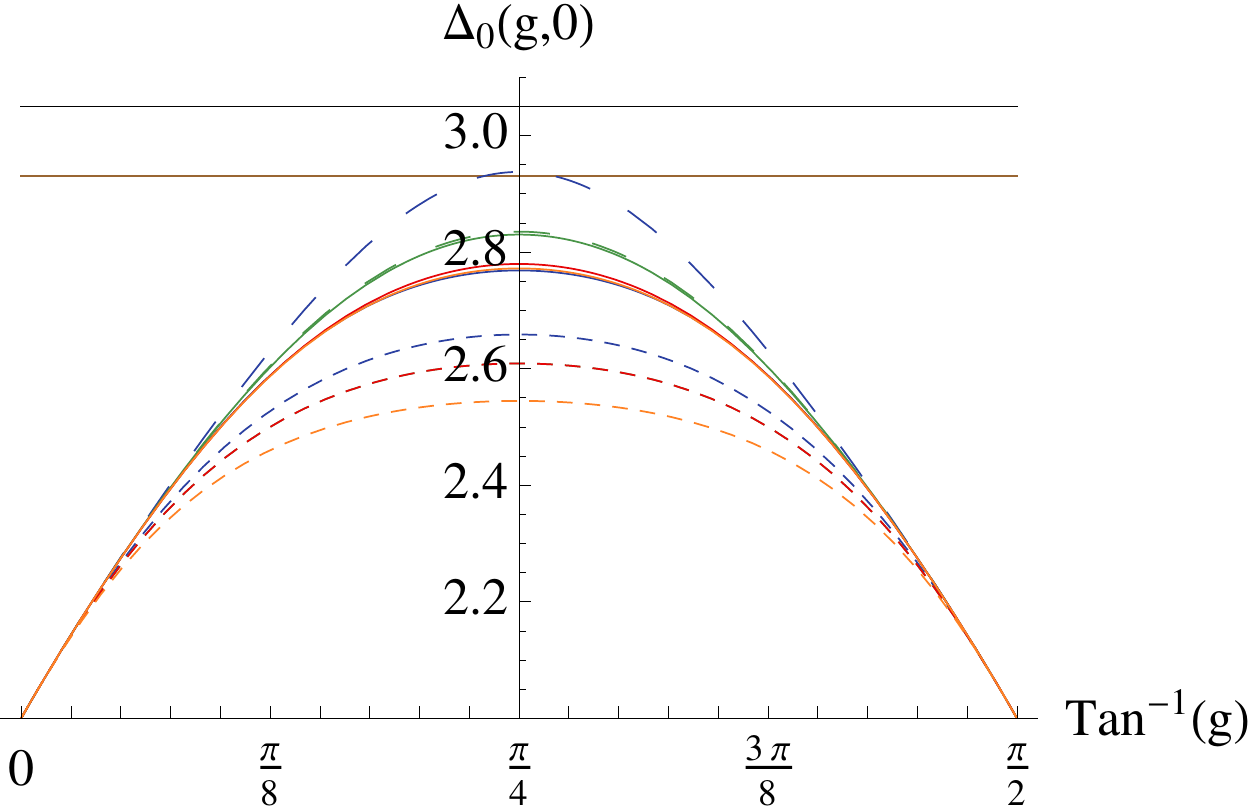}
}\vspace{8pt}
\hbox{
\epsfysize=4.85cm\epsfbox{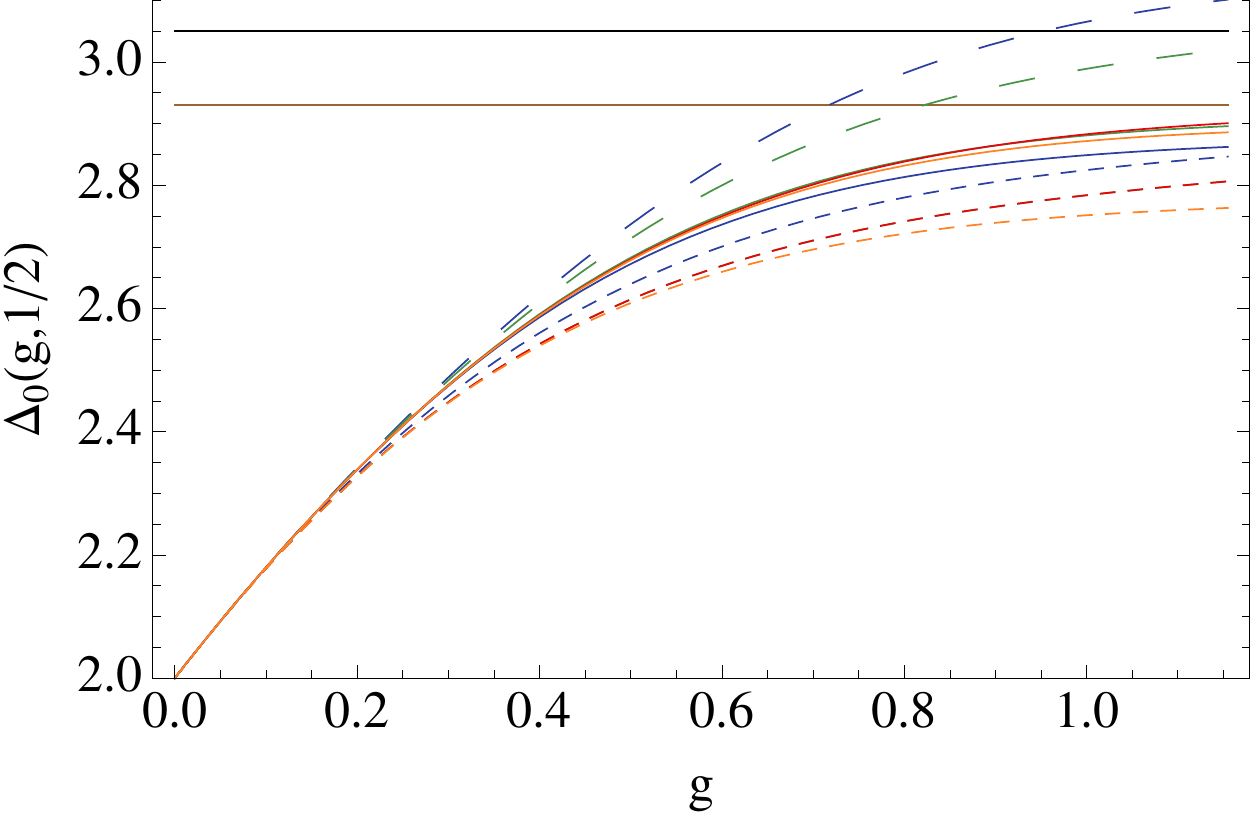}~~
\epsfysize=4.85cm\epsfbox{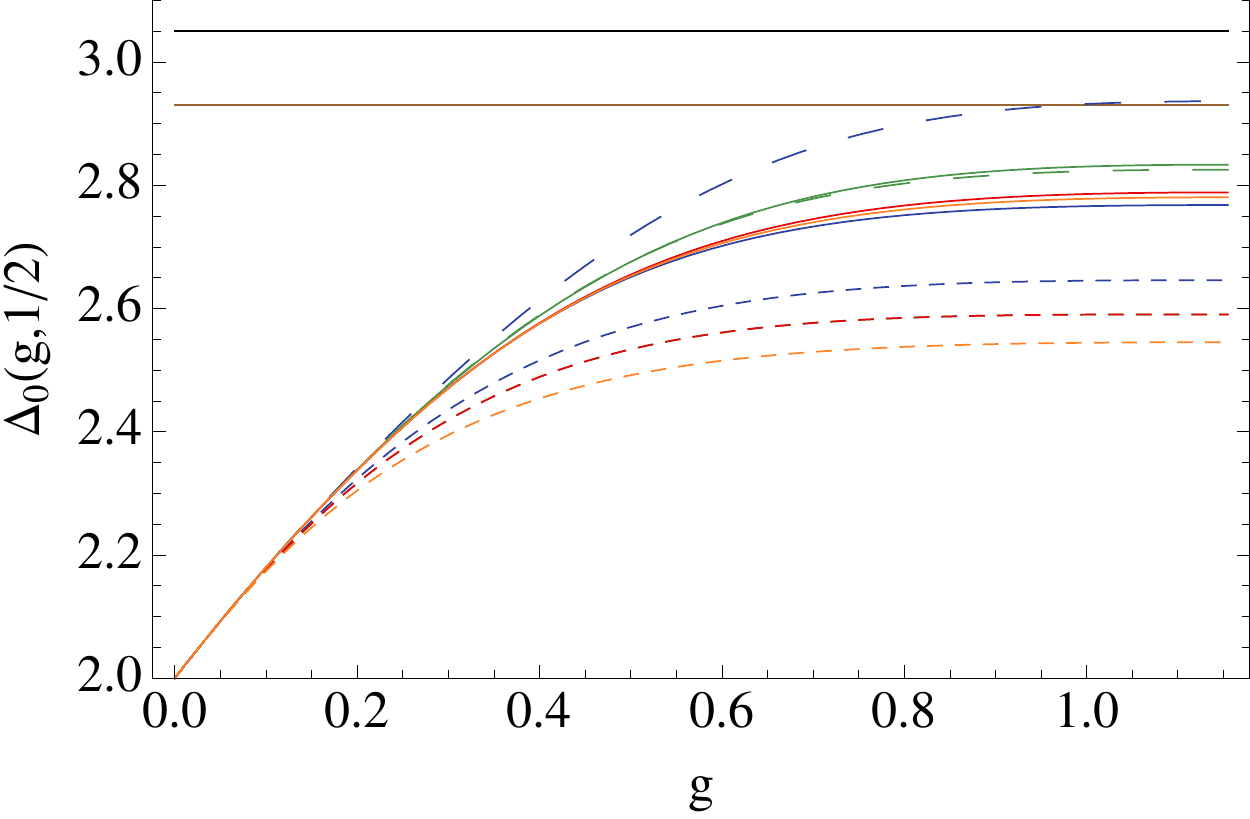}
}
\end{center}
\vspace{-20pt}
\caption{Interpolations of the Konishi anomalous dimensions for gauge group $SU(2)$. The different plots depict the results of the (left) $\BBZ_2$ invariant and (right) $\BBZ_3$ invariant resummation schemes, evaluated as a function of $g$ with (top) $\theta=0$ and (bottom) $\theta=\pi$. We show interpolations defined using (short-dashed) two loops, (long-dashed) three loops, and (solid) four loops in perturbation theory. Red and orange lines correspond to Pad\'e approximants with integral and half-integral powers, respectively. Blue and green lines represent FPP interpolations with integral and half-integral powers. 
As described at the end of appendix \ref{sa}, some of these graphs coincide. 
The two horizontal lines correspond to the upper bound (top line) and the best estimate based on a corner value (bottom line) obtained from the numerical bootstrap results of \cite{1304.1803}. See \S\ref{s5} for a more detailed
description of these bounds.
\label{fig:Spin0SU2}}
\end{figure}}

\def\spinzeroSUthree{
\begin{figure}[ht]
\begin{center}
\hbox{
\epsfysize=4.85cm\epsfbox{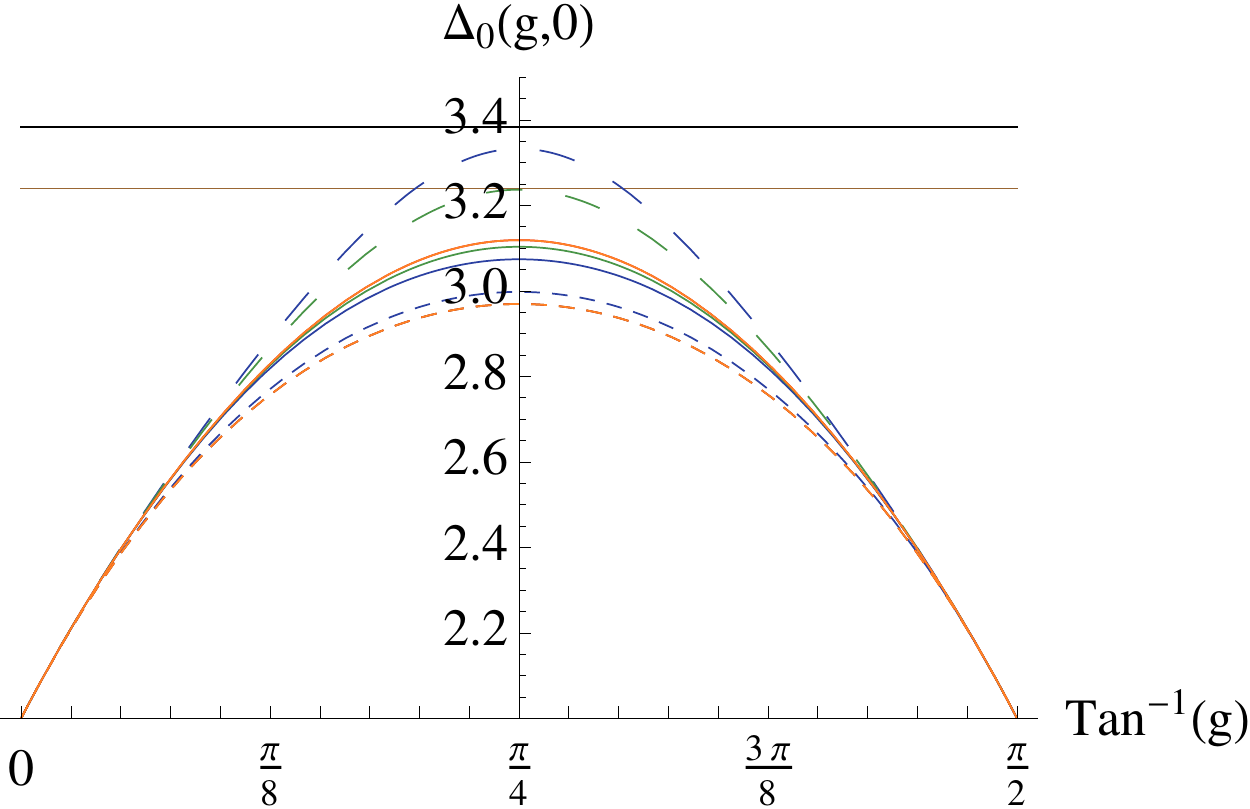}~~
\epsfysize=4.85cm\epsfbox{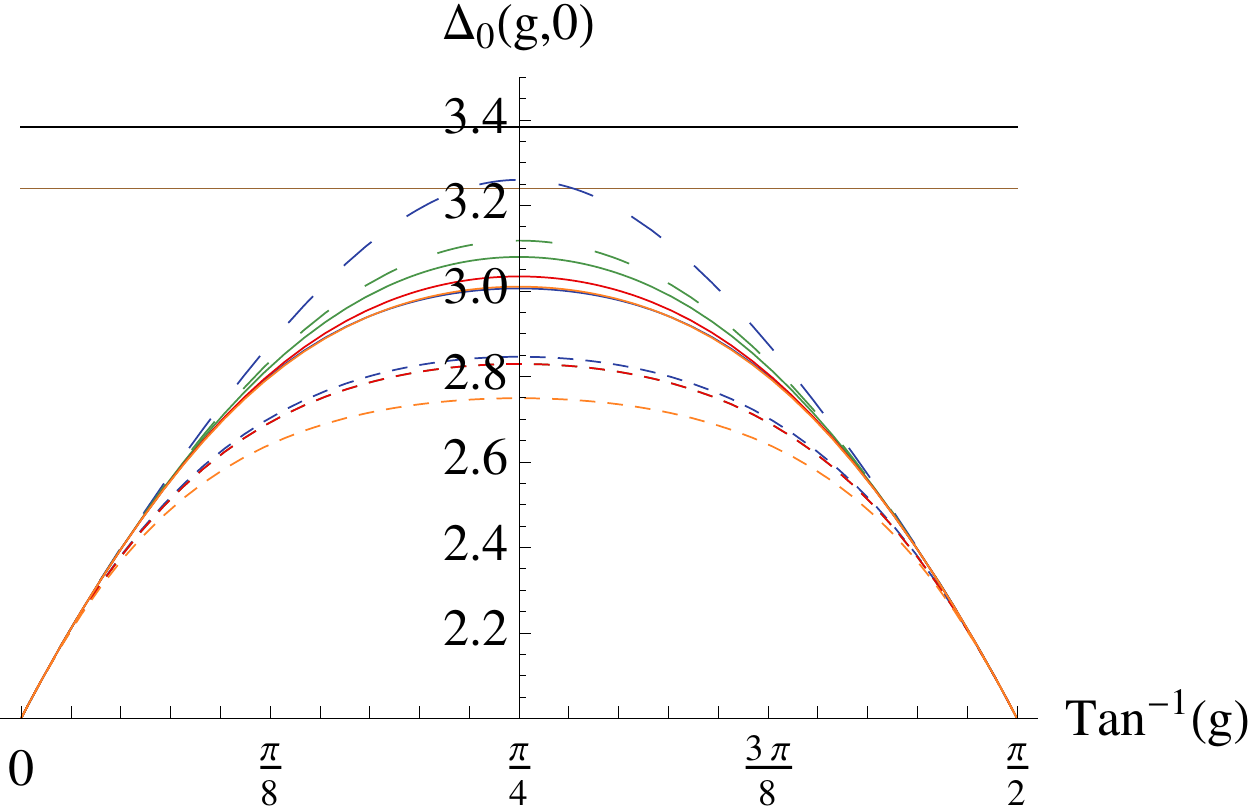}
}
\vspace{8pt}
\hbox{
\epsfysize=4.85cm\epsfbox{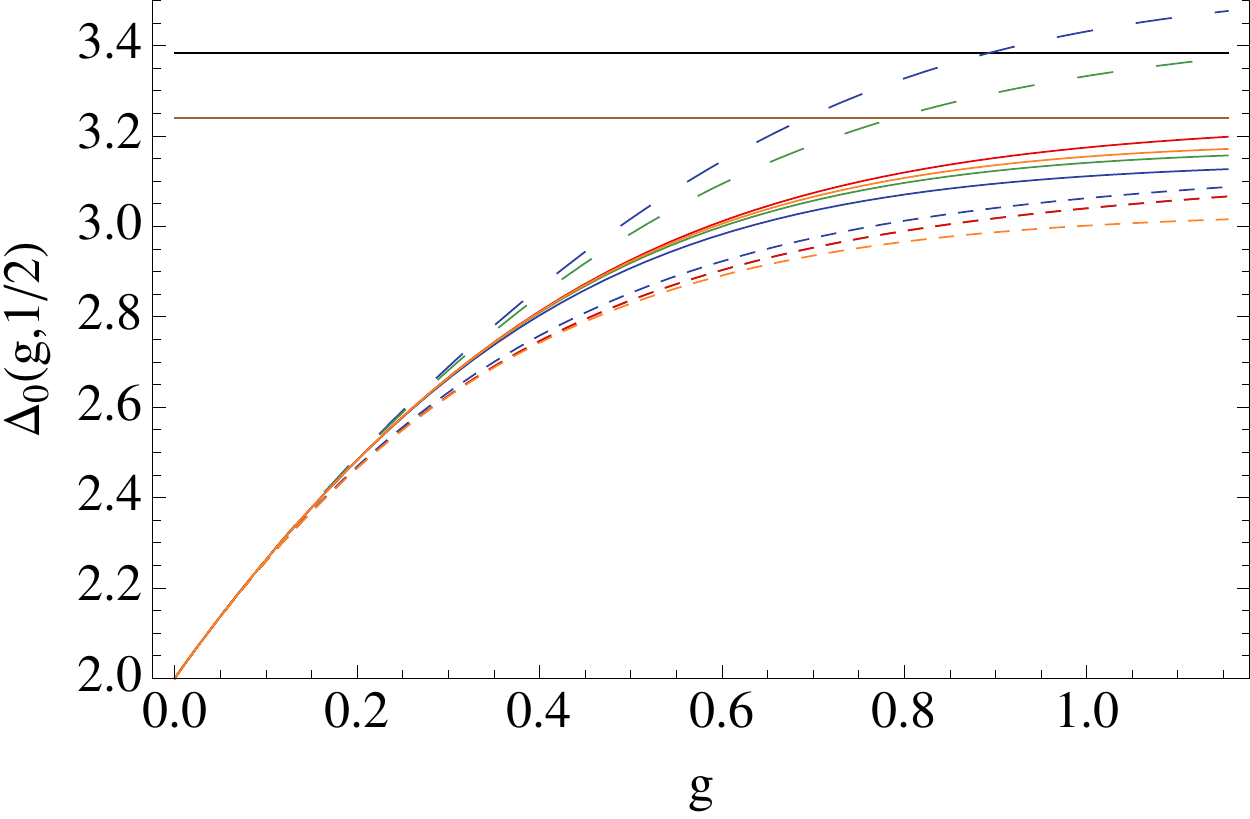}~~
\epsfysize=4.85cm\epsfbox{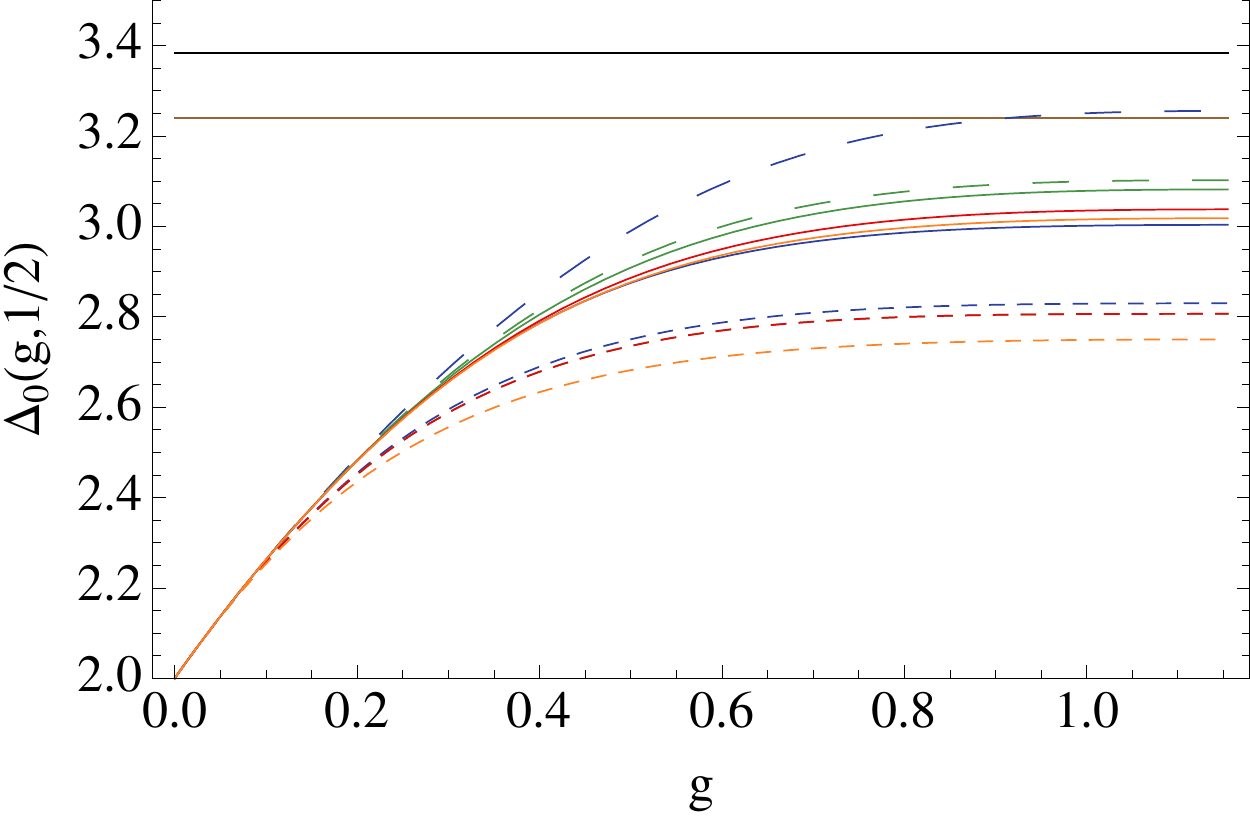}
}
\end{center}
\vspace{-20pt}
\caption{Interpolations of the Konishi anomalous dimensions for gauge group $SU(3)$. The different plots depict the results of the (left) $\BBZ_2$ invariant and (right) $\BBZ_3$ invariant resummation schemes, evaluated as a function of $g$ with (top) $\theta=0$ and (bottom) $\theta=\pi$.
\label{fig:Spin0SU3}}
\end{figure}}

\def\spinzeroSUfour{
\begin{figure}[ht]
\begin{center}
\hbox{
\epsfysize=4.85cm\epsfbox{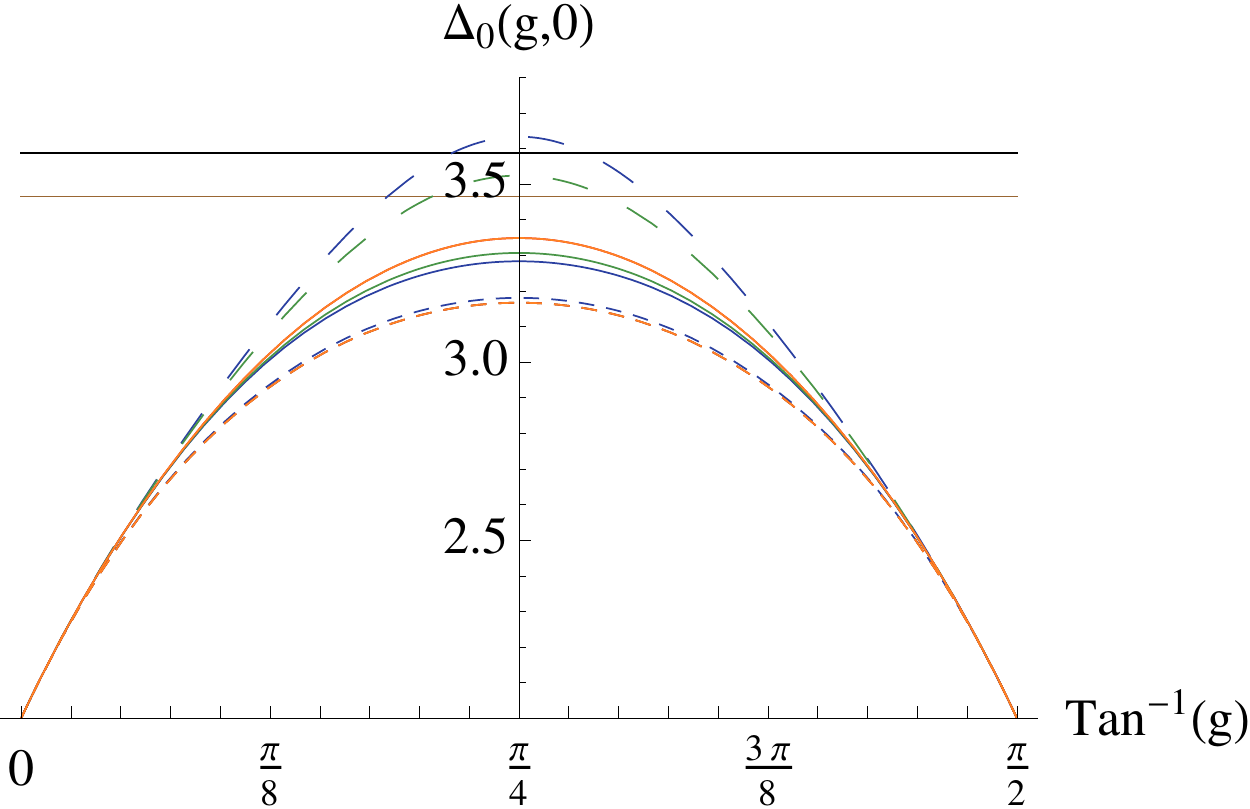}~~
\epsfysize=4.85cm\epsfbox{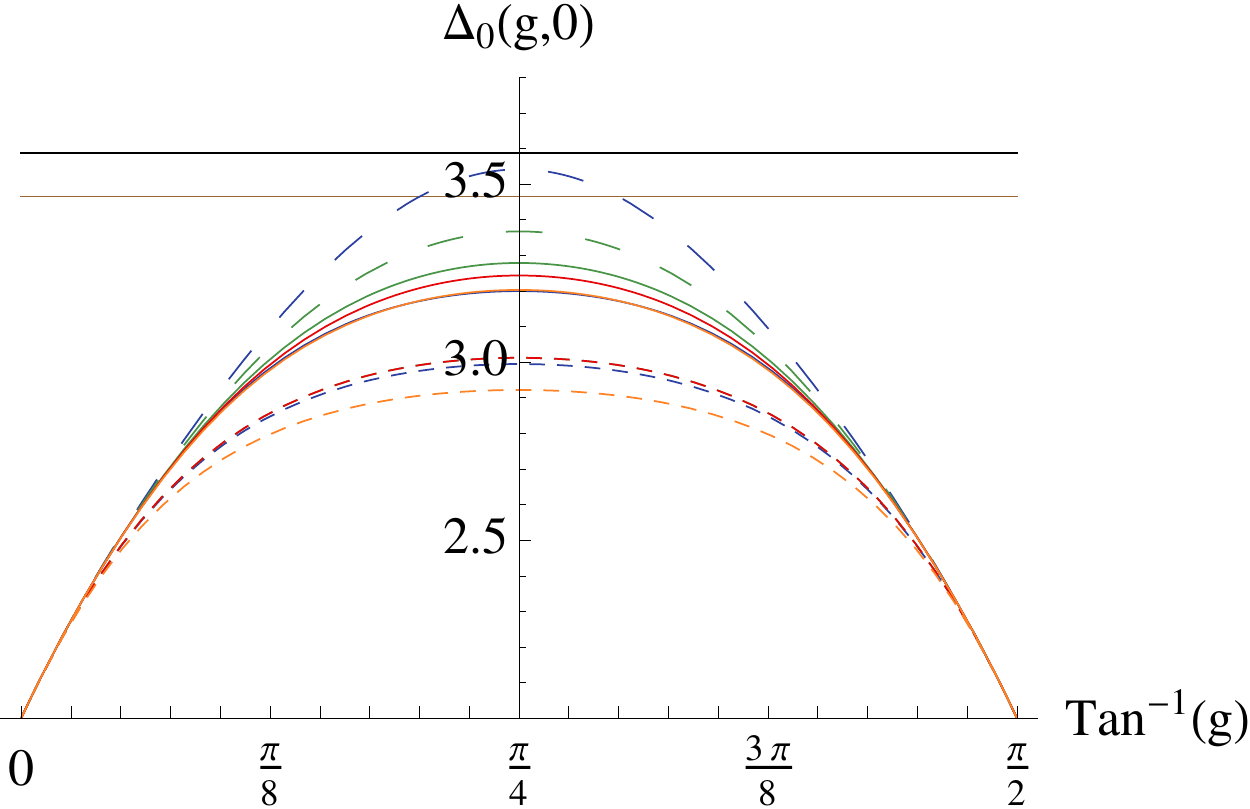}
}
\vspace{8pt}
\hbox{
\epsfysize=4.85cm\epsfbox{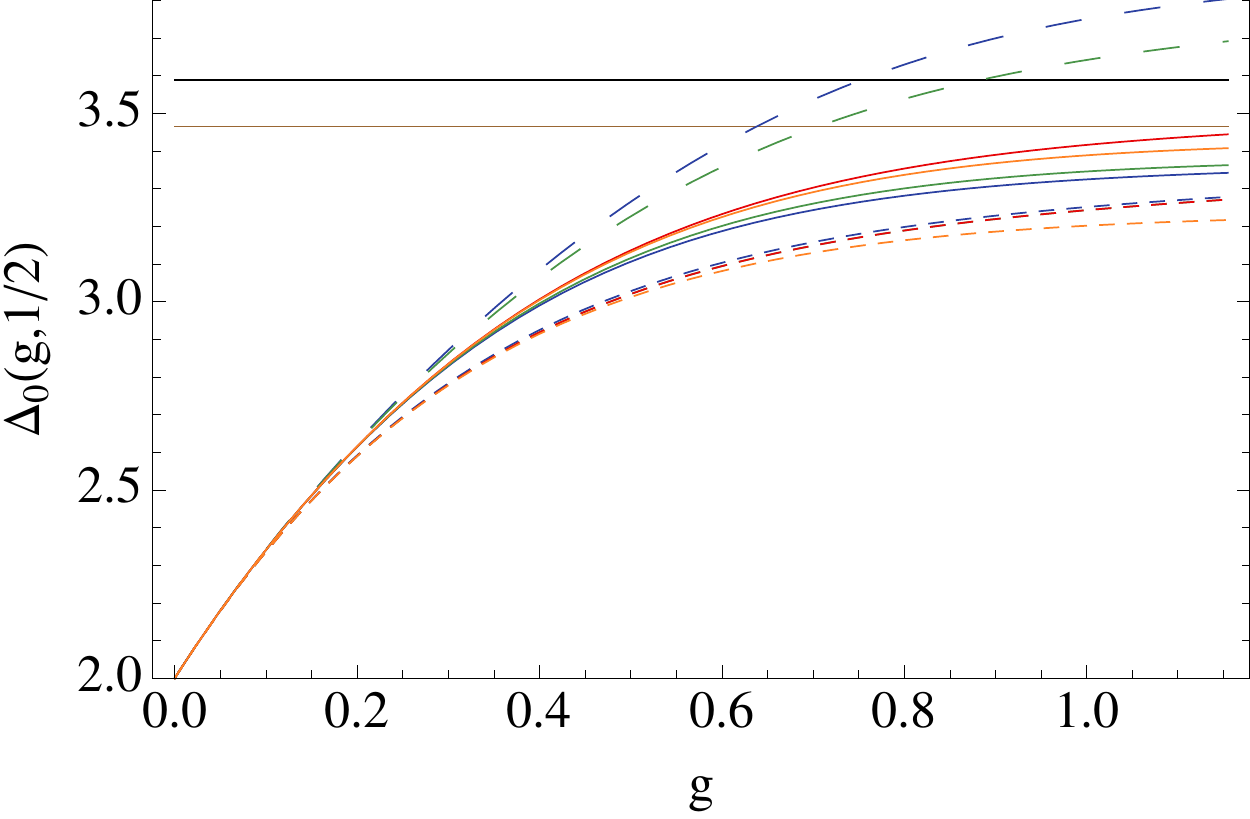}~~
\epsfysize=4.85cm\epsfbox{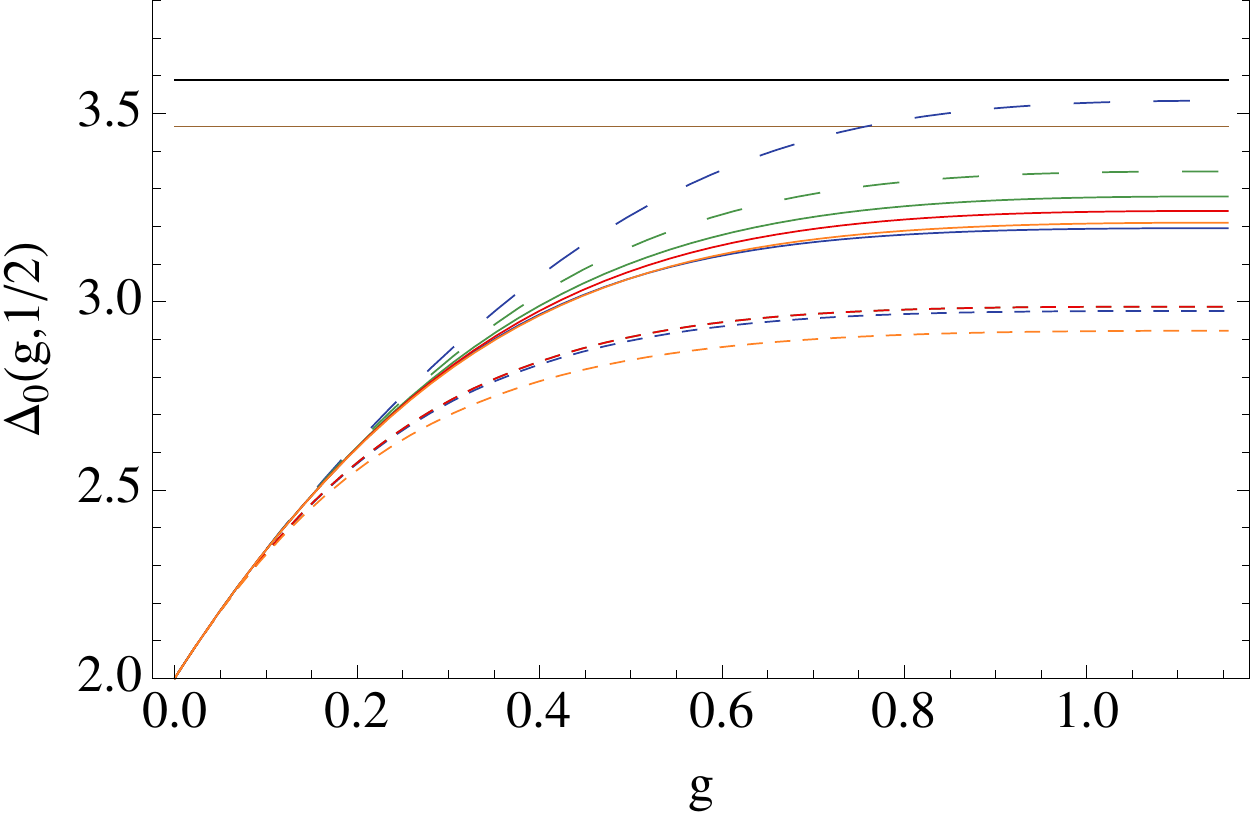}
}
\end{center}
\vspace{-20pt}
\caption{Interpolations of the Konishi anomalous dimensions for gauge group $SU(4)$. The different plots depict the results of the (left) $\BBZ_2$ invariant and (right) $\BBZ_3$ invariant resummation schemes, evaluated as a function of $g$ with (top) $\theta=0$ and (bottom) $\theta=\pi$.
\label{fig:Spin0SU4}}
\end{figure}}

\def\spintwoSUtwo{
\begin{figure}%[ht]
\begin{center}
\hbox{
\epsfysize=4.85cm\epsfbox{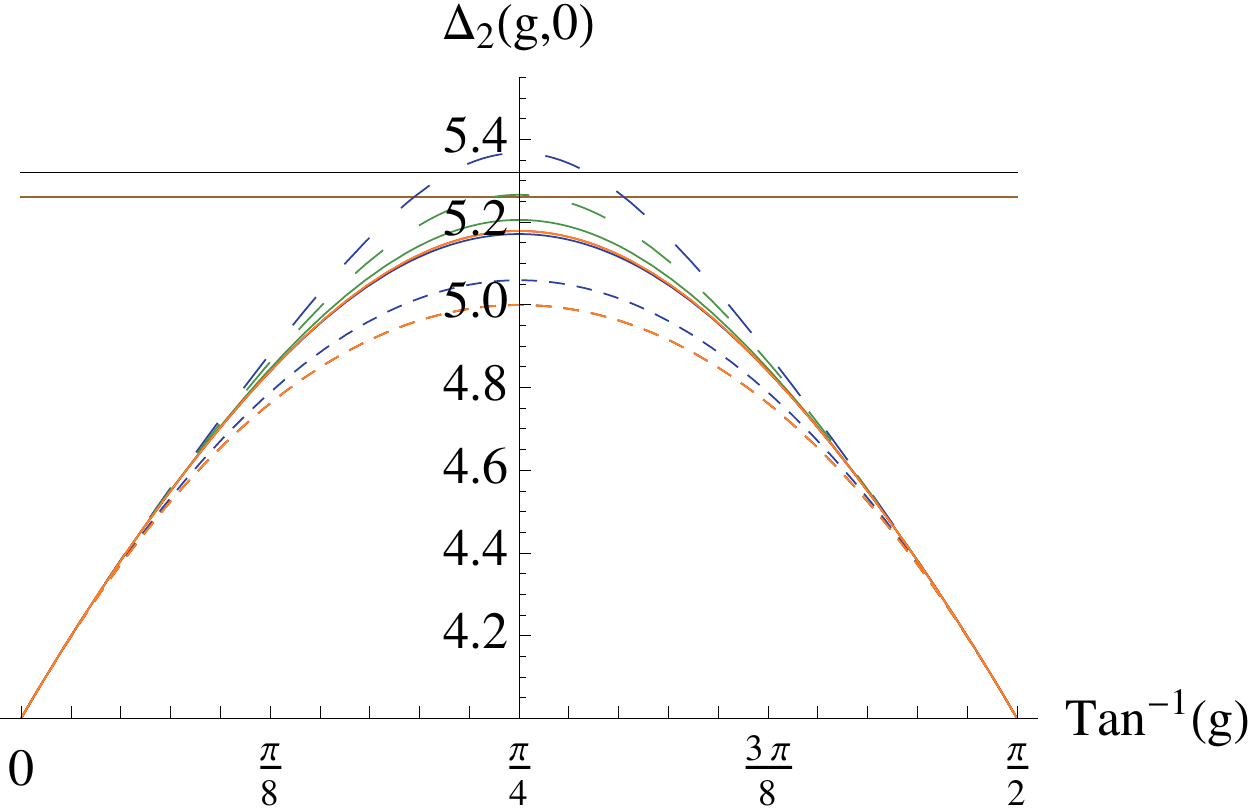}~~
\epsfysize=4.85cm\epsfbox{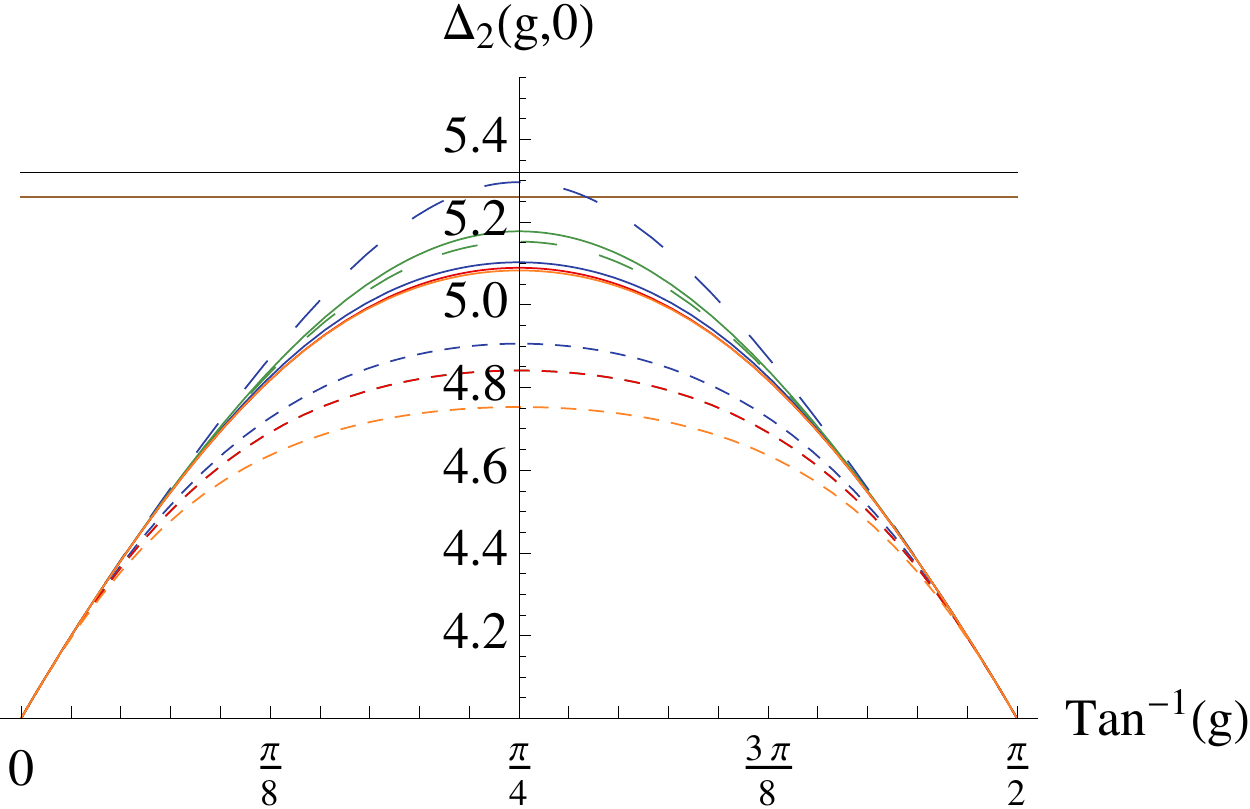}
}\vspace{8pt}
\hbox{
\epsfysize=4.85cm\epsfbox{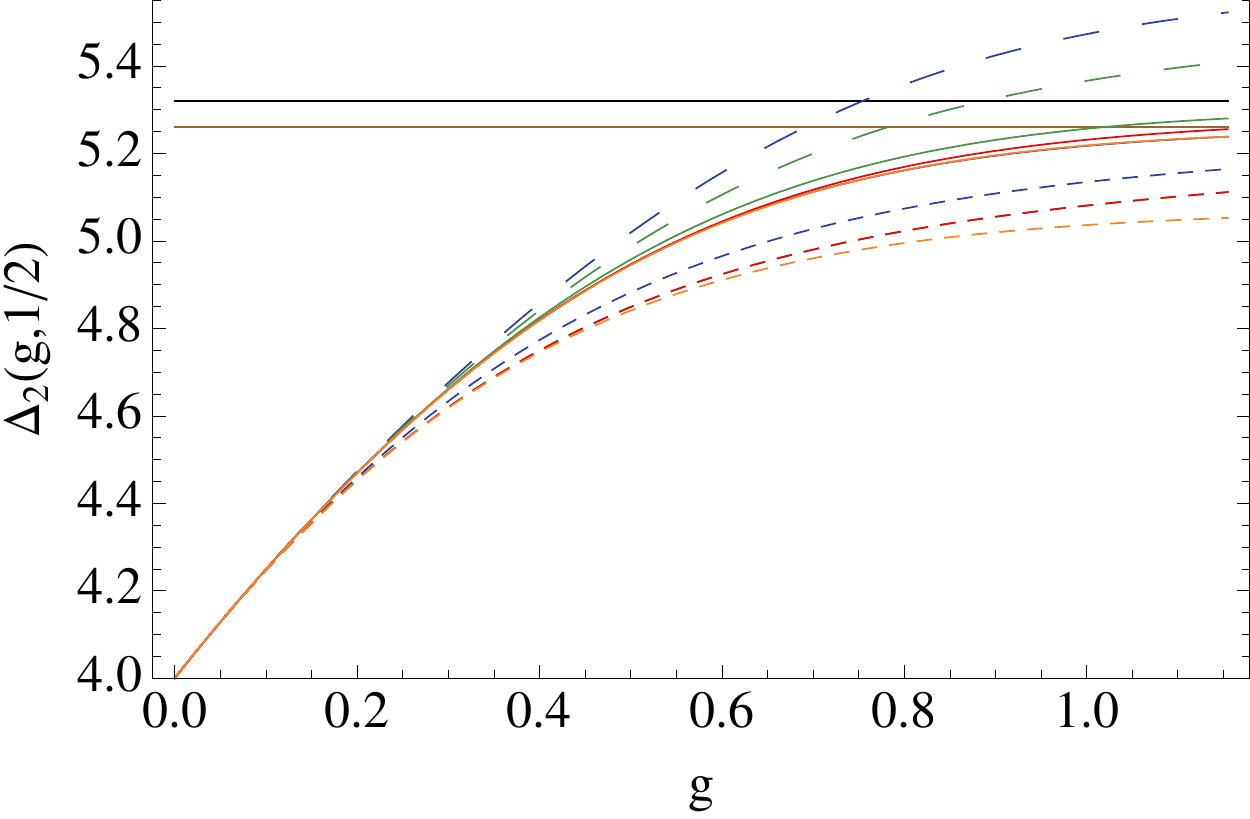}~~
\epsfysize=4.85cm\epsfbox{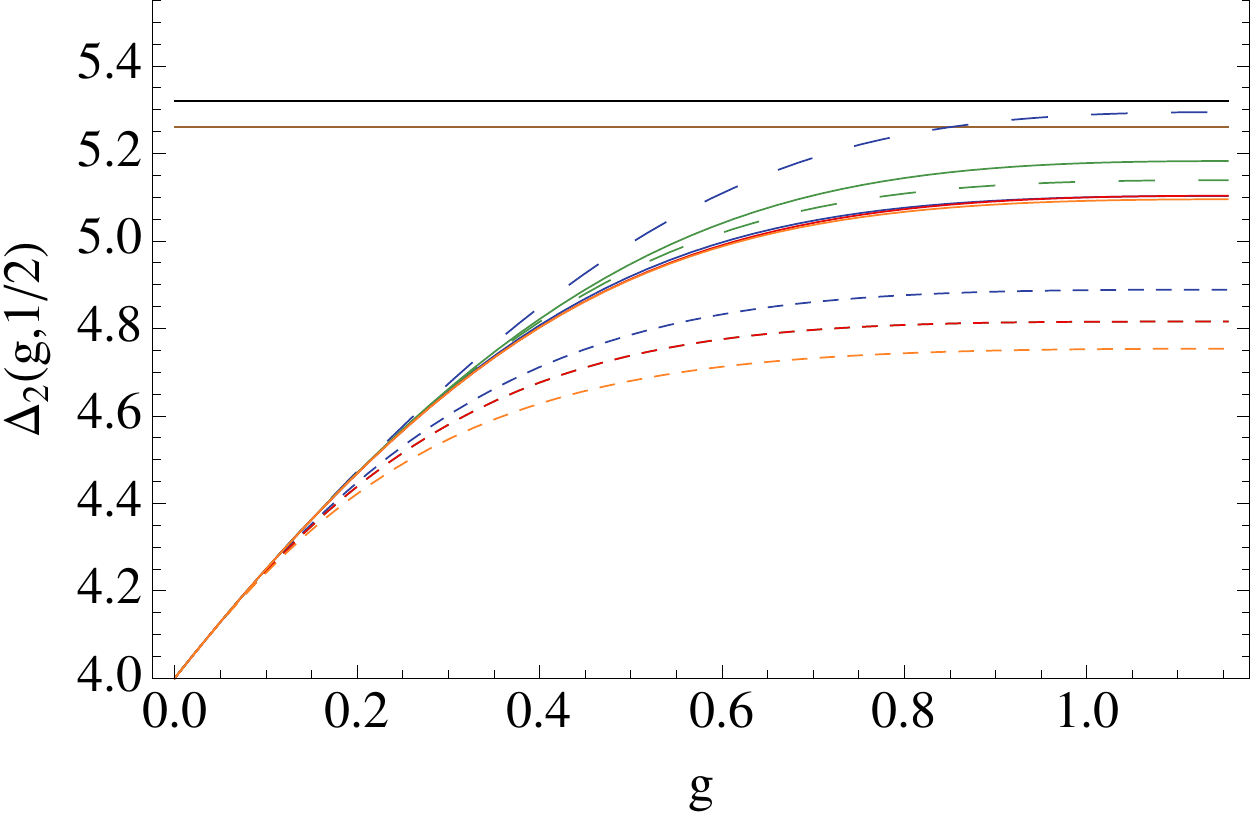}
}
\end{center}
\vspace{-20pt}
\caption{Interpolations of the spin two anomalous dimensions for gauge group $SU(2)$. The different plots depict the results of the (left) $\BBZ_2$ invariant and (right) $\BBZ_3$ invariant resummation schemes, evaluated as a function of $g$ with (top) $\theta=0$ and (bottom) $\theta=\pi$.
\label{fig:Spin2SU2}}
\end{figure}}

\def\spintwoSUthree{
\begin{figure}%[hb]
\begin{center}
\hbox{
\epsfysize=4.85cm\epsfbox{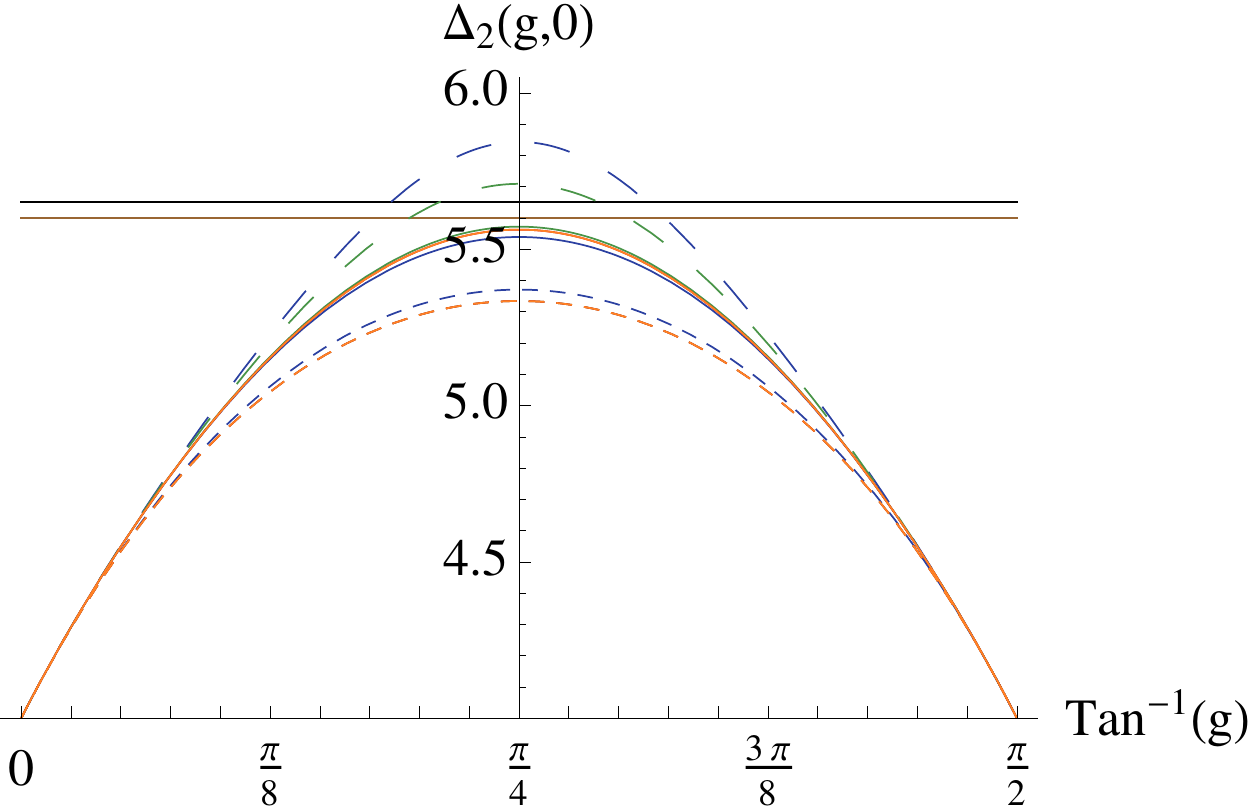}~~
\epsfysize=4.85cm\epsfbox{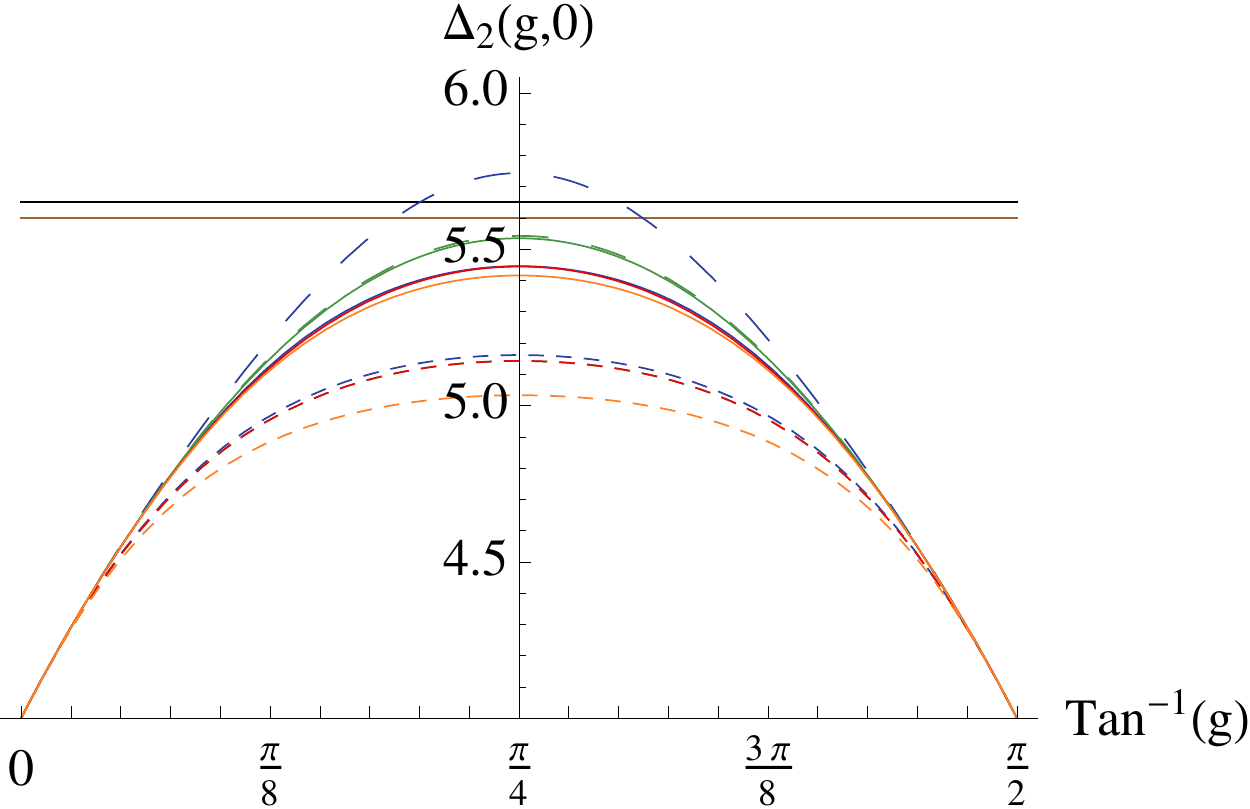}
}
\vspace{8pt}
\hbox{
\epsfysize=4.85cm\epsfbox{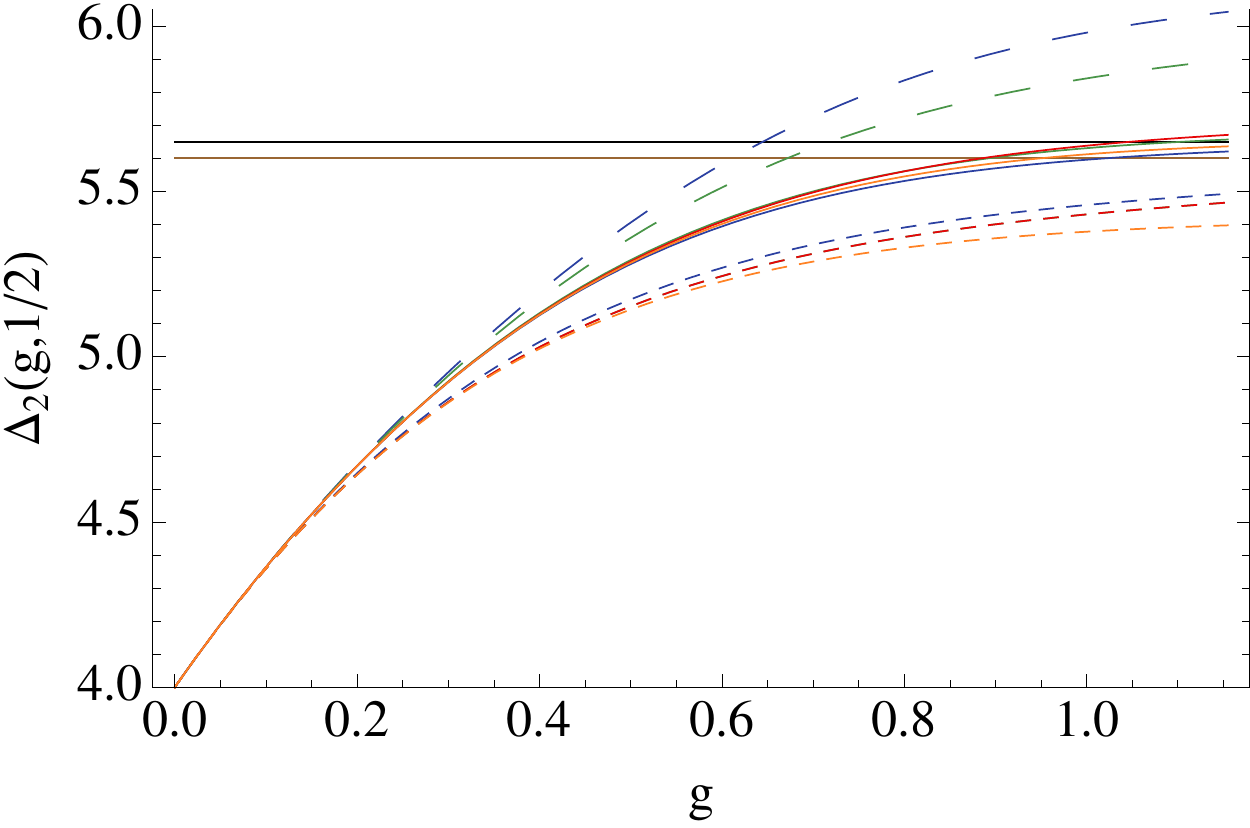}~~
\epsfysize=4.85cm\epsfbox{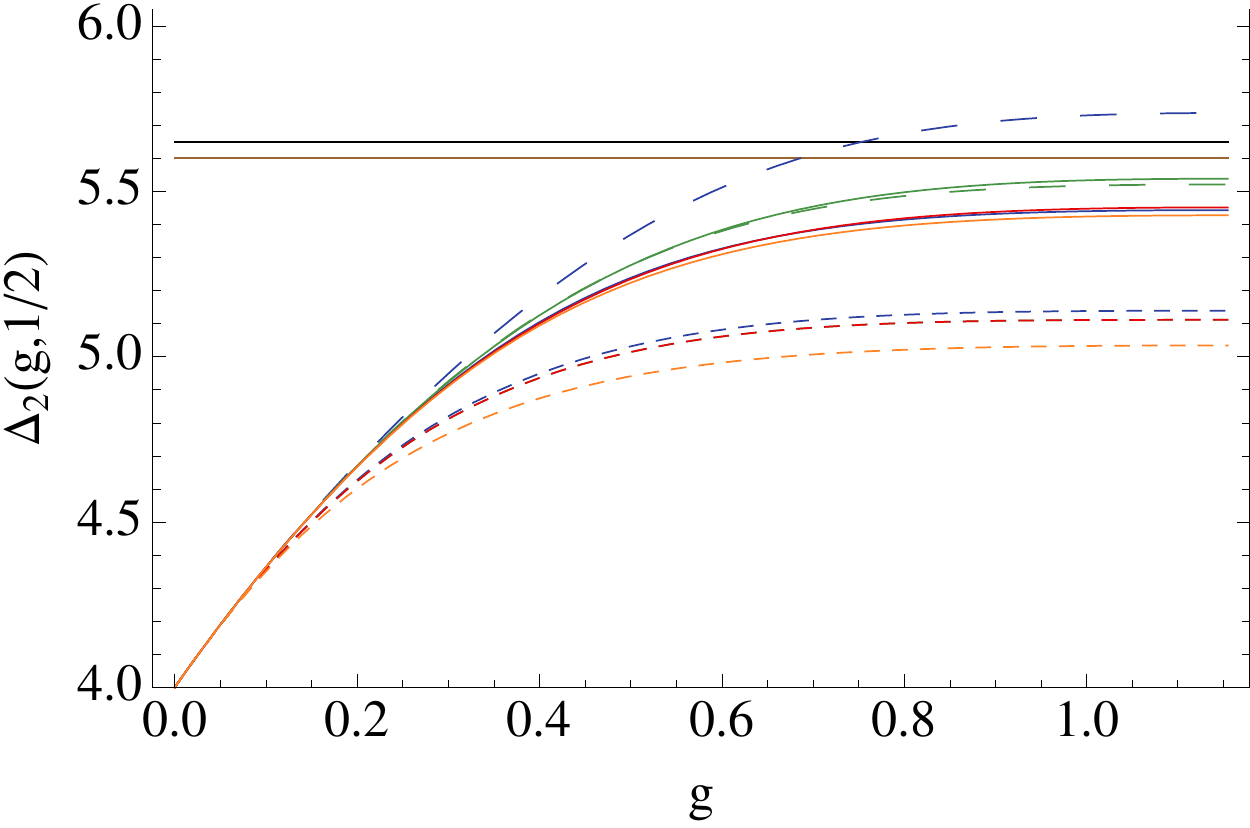}
}
\end{center}
\vspace{-20pt}
\caption{Interpolations of the spin two anomalous dimensions for gauge group $SU(3)$. The different plots depict the results of the (left) $\BBZ_2$ invariant and (right) $\BBZ_3$ invariant resummation schemes, evaluated as a function of $g$ with (top) $\theta=0$ and (bottom) $\theta=\pi$.
\label{fig:Spin2SU3}}
\end{figure}}

\def\spintwoSUfour{
\begin{figure}%[ht]
\begin{center}
\hbox{
\epsfysize=4.85cm\epsfbox{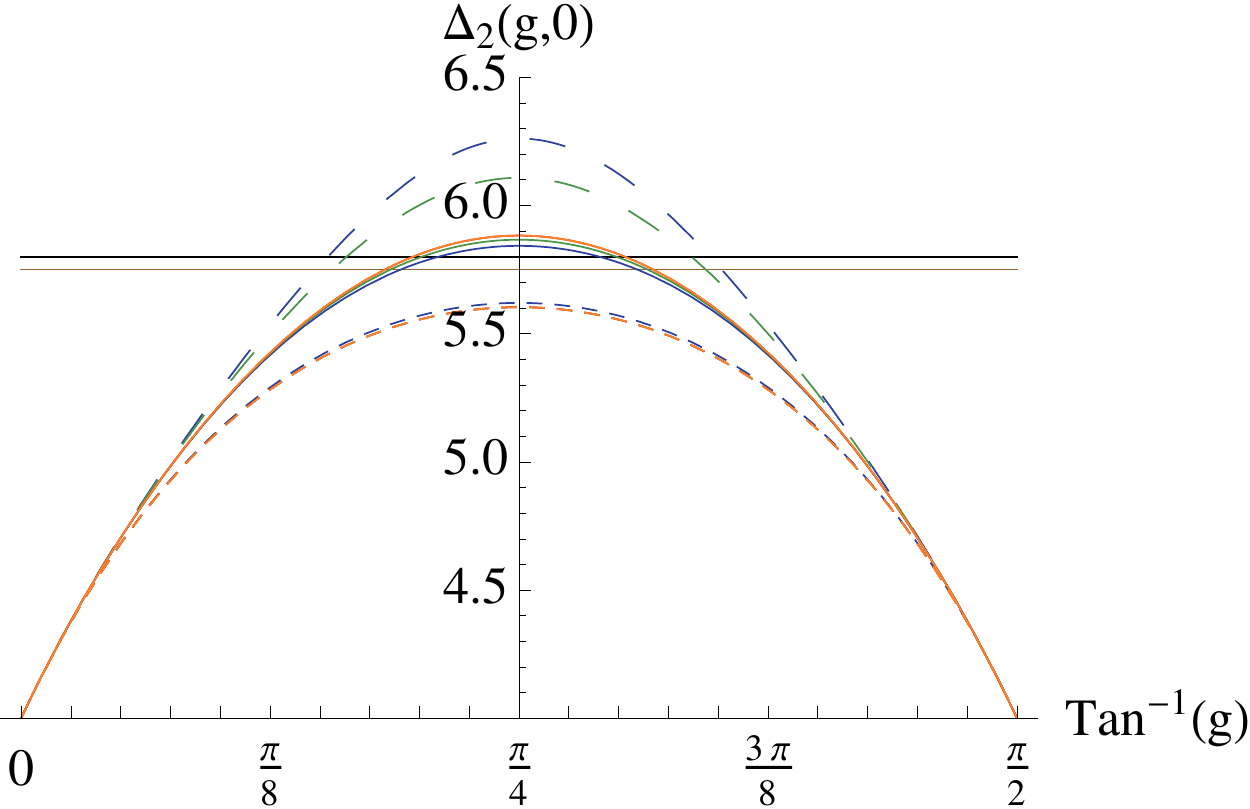}~~
\epsfysize=4.85cm\epsfbox{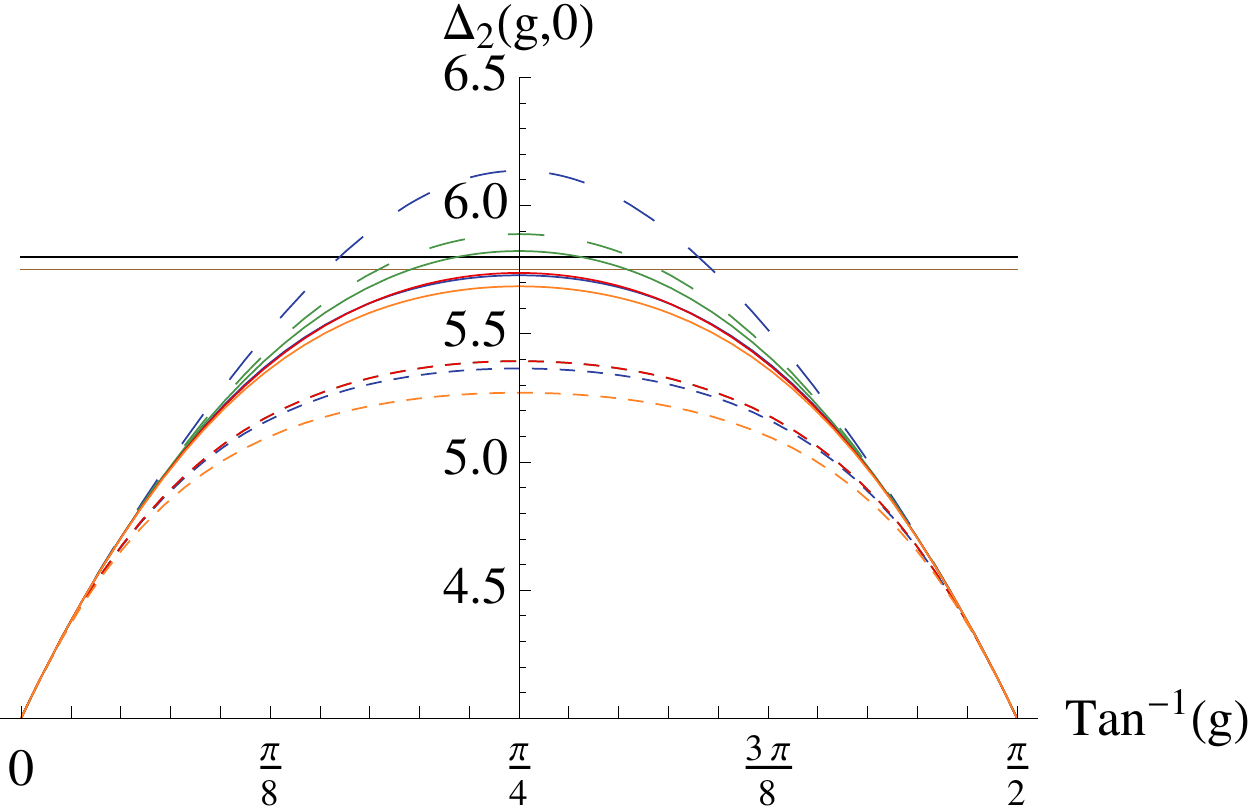}
}
\vspace{8pt}
\hbox{
\epsfysize=4.85cm\epsfbox{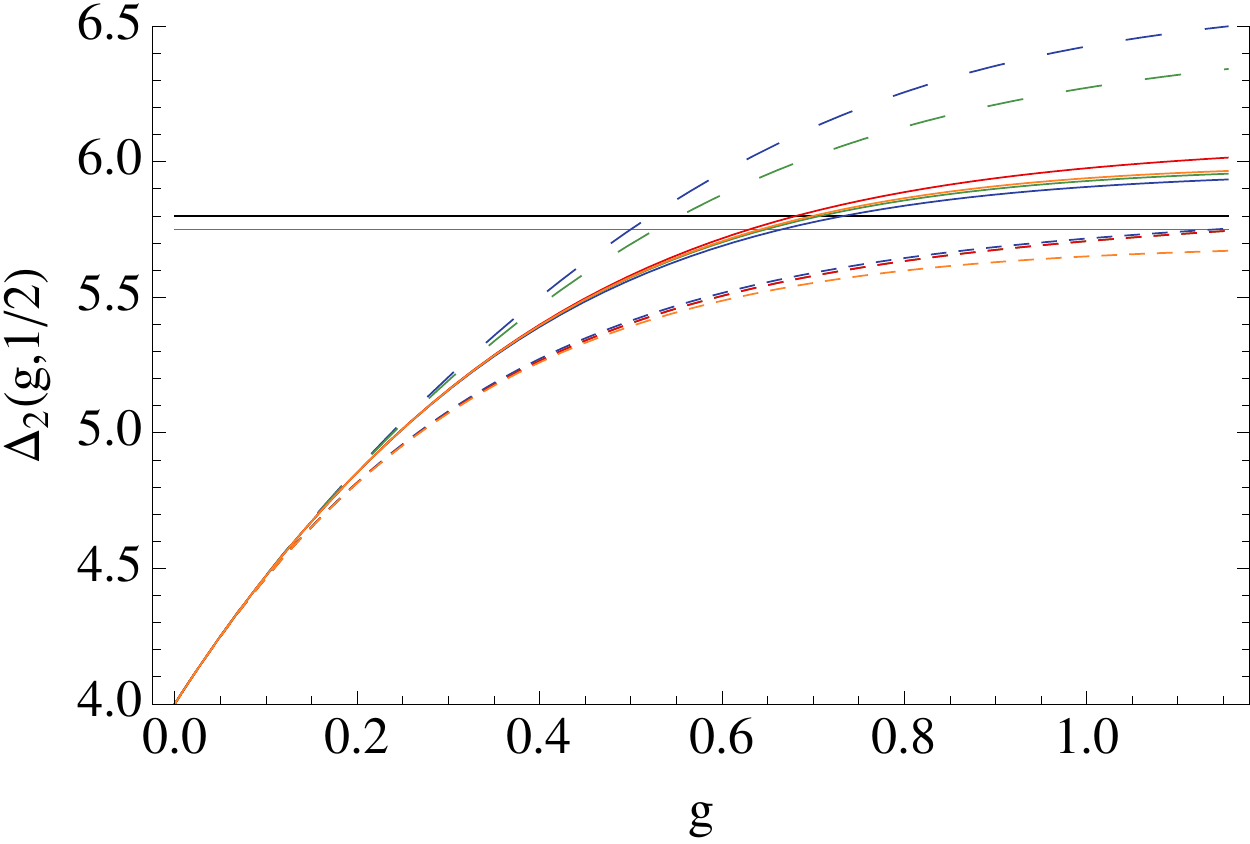}~~
\epsfysize=4.85cm\epsfbox{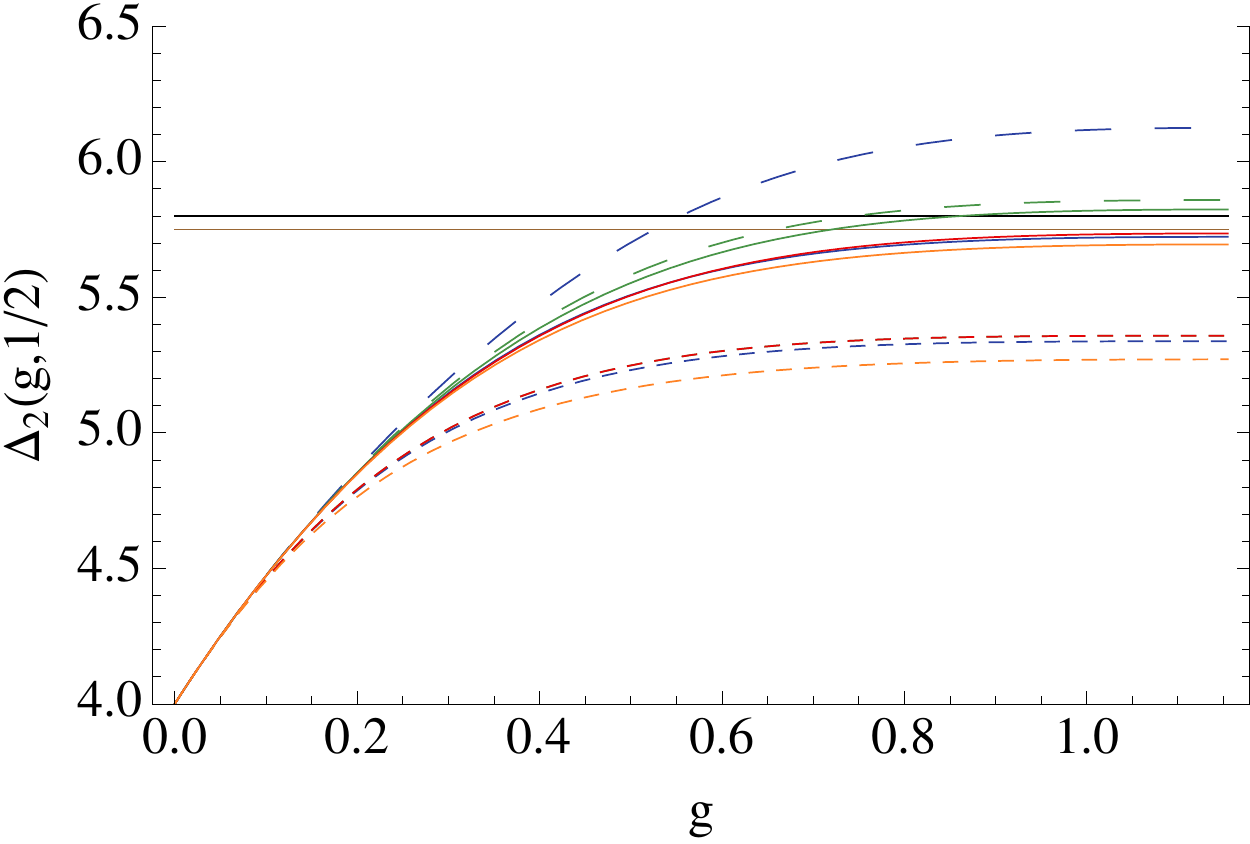}
}
\end{center}
\vspace{-20pt}
\caption{Interpolations of the spin two anomalous dimensions for gauge group $SU(4)$. The different plots depict the results of the (left) $\BBZ_2$ invariant and (right) $\BBZ_3$ invariant resummation schemes, evaluated as a function of $g$ with (top) $\theta=0$ and (bottom) $\theta=\pi$.
\label{fig:Spin2SU4}}
\end{figure}}

\def\bootstrapCompare{
\begin{figure}%[ht]
\begin{center}
\hbox{
\epsfysize=4.85cm\epsfbox{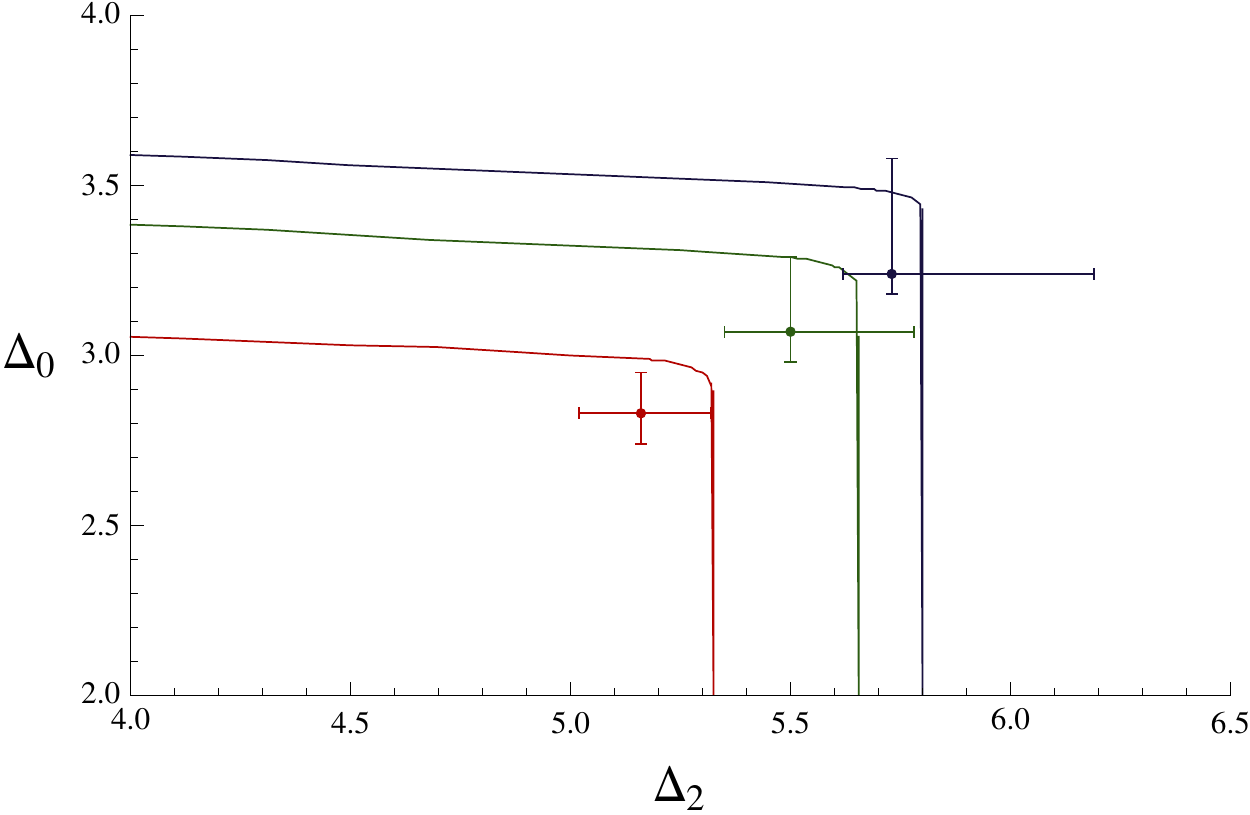}~~
\epsfysize=4.85cm\epsfbox{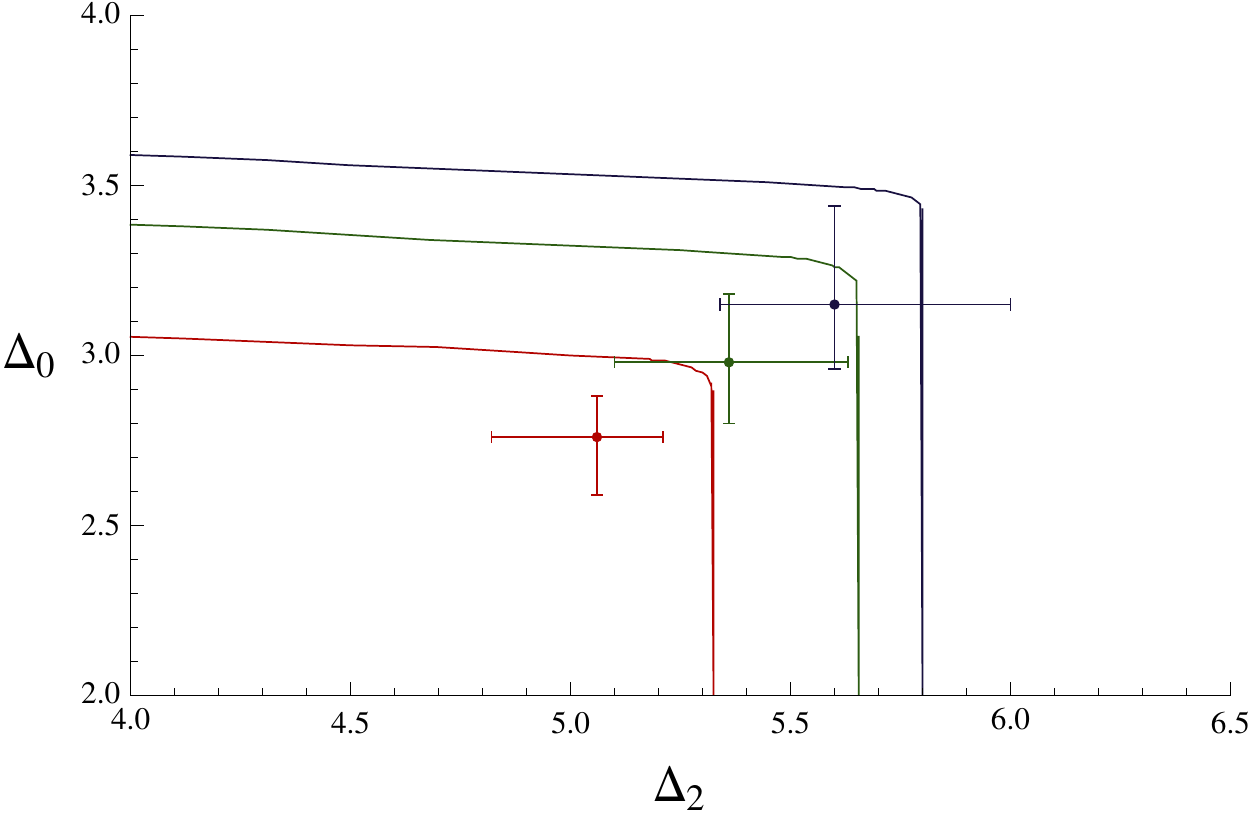}
}
\end{center}
\vspace{-20pt}
\caption{These figures juxtapose the results of the (left) $\BBZ_2$ invariant resummations evaluated at $\tau_2$ and (right) $\BBZ_3$ invariant interpolations evaluated at $\tau_3$ with the bounds obtained in \cite{1304.1803} for dimensions of spin zero and spin two operators. The interpolation results 
are all compatible with the bounds, and moreover can be interpreted as supporting the conjecture that the optimal version of such bounds is saturated at one of these points.\label{fig:comparison}}
\end{figure}}

\def\figureconformal{
\begin{figure}%[ht]
\begin{center}
\hbox{
\epsfysize=4.9cm\epsfbox{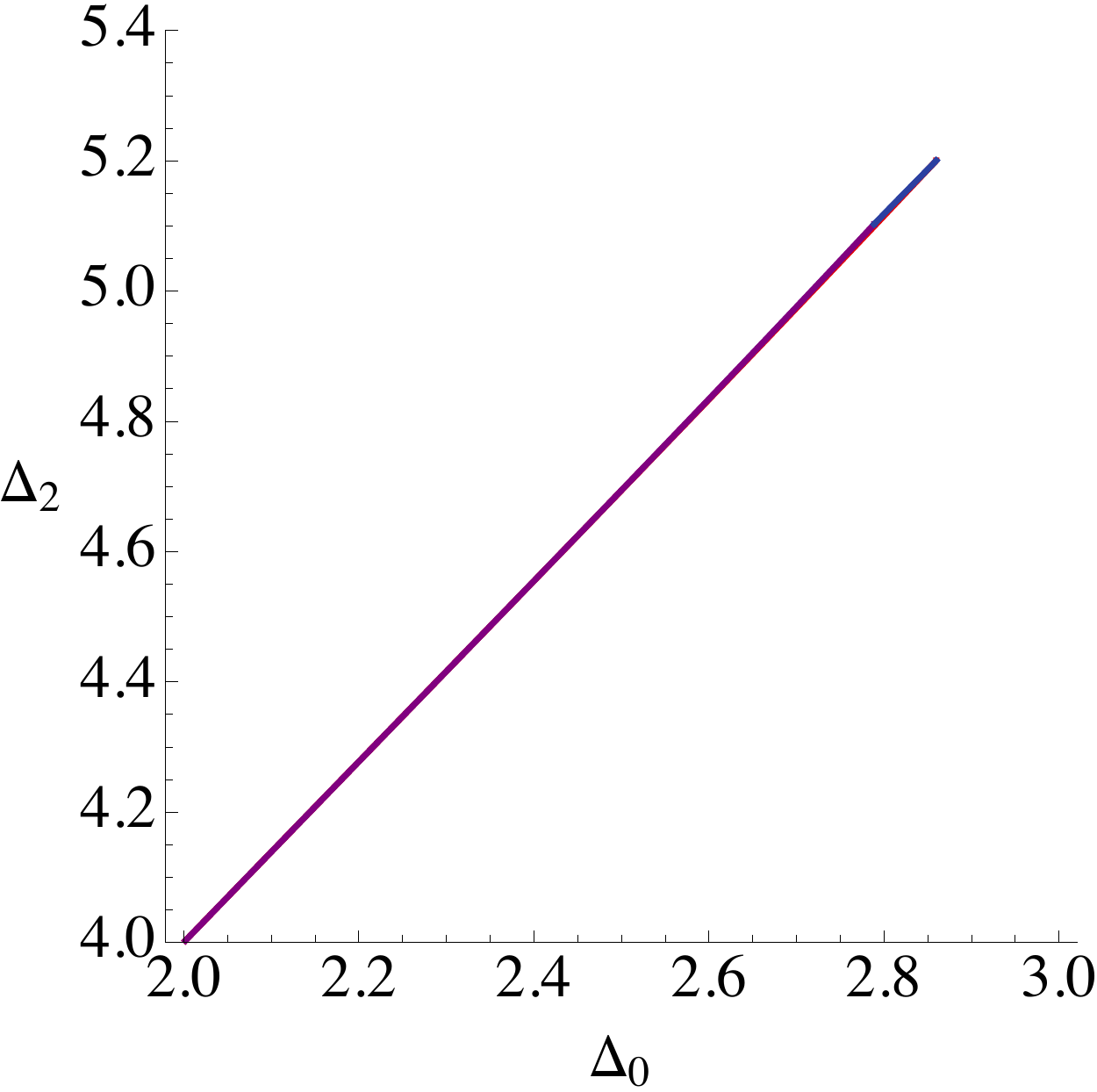}
\epsfysize=4.9cm\epsfbox{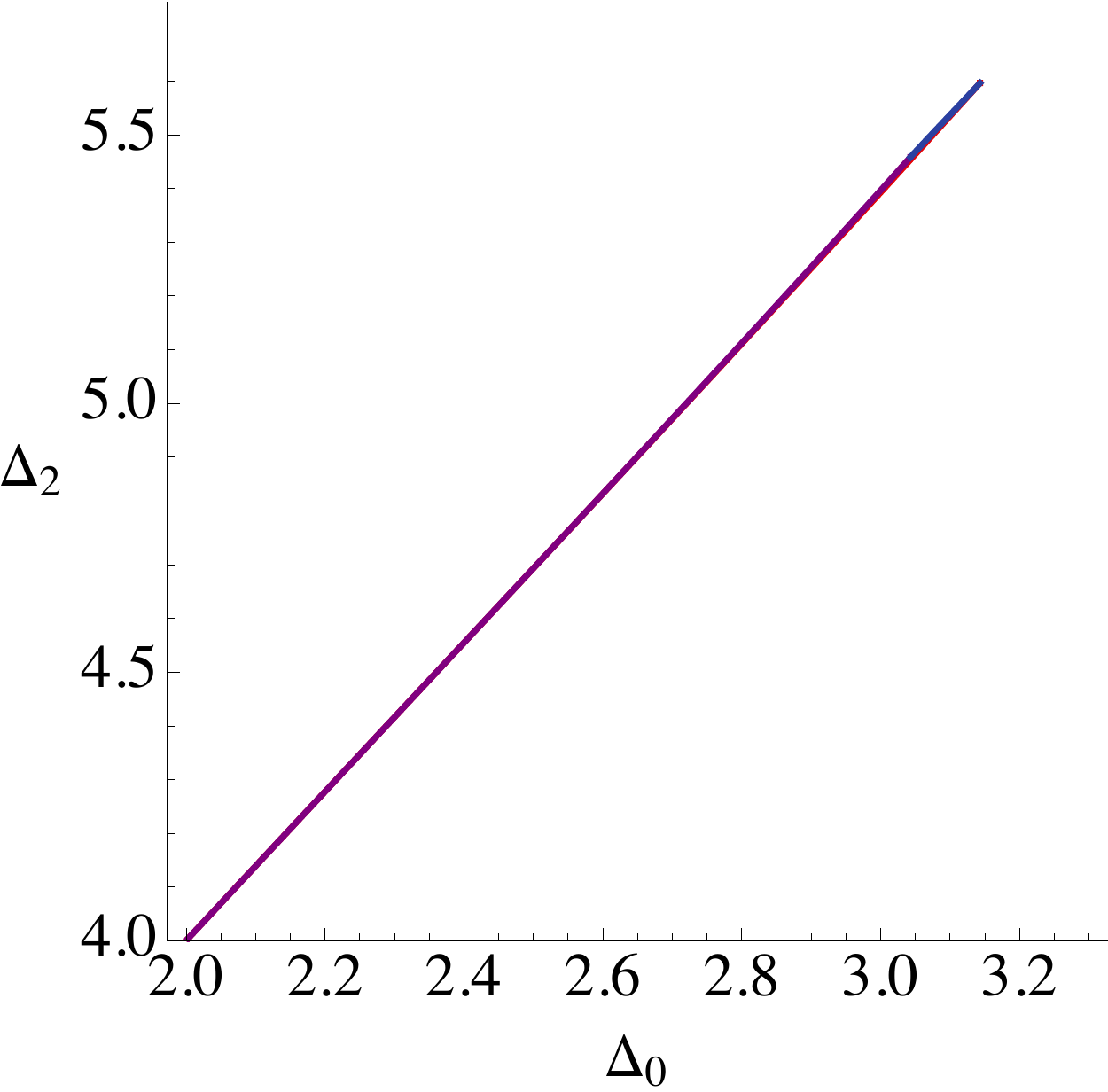}
\epsfysize=4.9cm\epsfbox{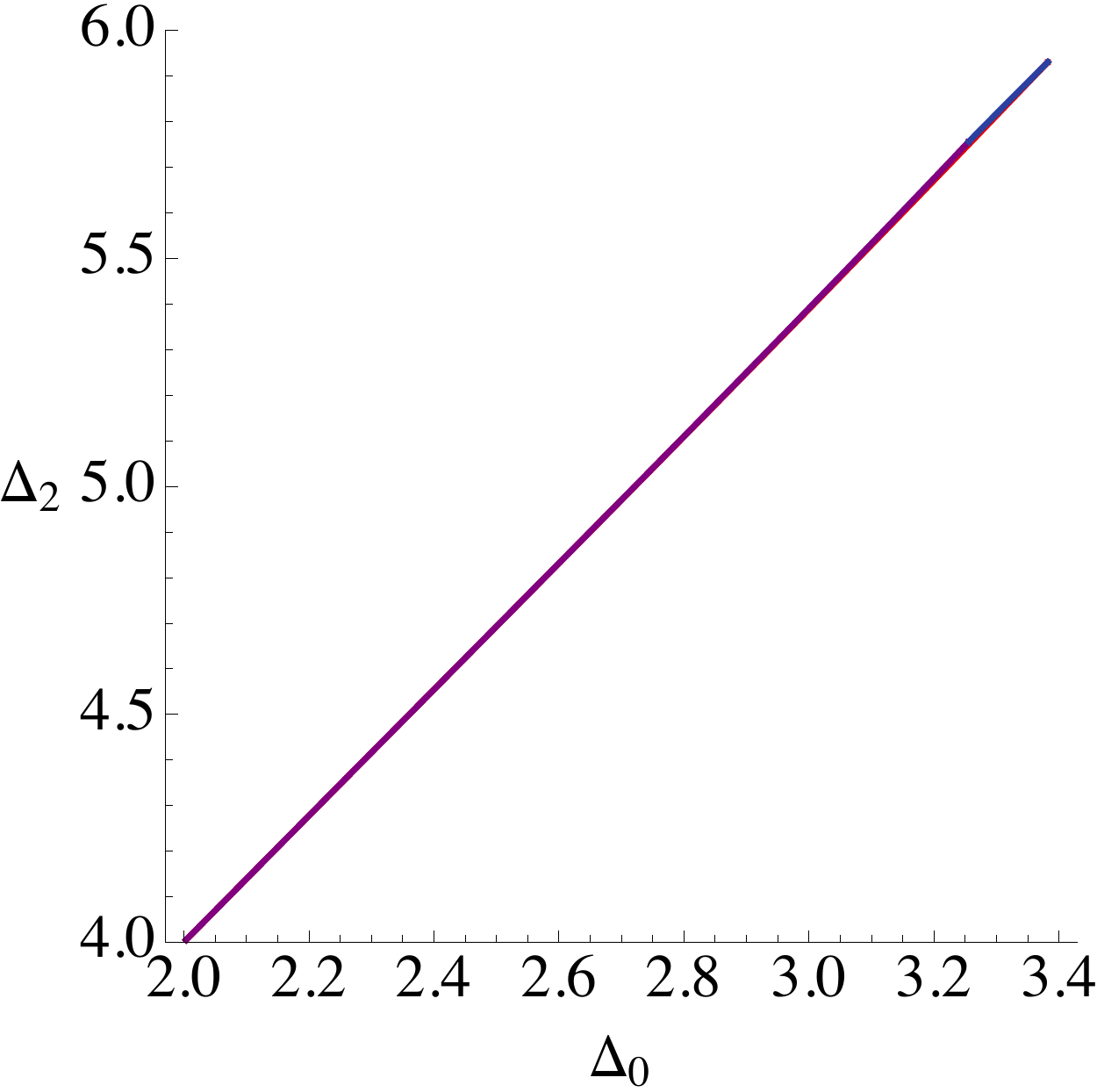}
}
\end{center}
\vspace{-20pt}
\caption{This figure shows the image of the conformal manifold in the
$(\Delta_0, \Delta_2)$ plane. The left graph is for $SU(2)$, the middle graph is for
$SU(3)$, and the right graph is for $SU(4)$. The red curve is the image of the
$\theta=0$ line, the purple curve is the image of the $\theta=\pi$ line and
the blue curve is the image of the circle at $|\tau|=1$. However the red curve
is practically invisible as it is hidden below the purple and the blue curves.
\label{fconformal}}
\end{figure}
}

%% file: resummation_tables.tex
\def\spinZeroTwoTableNoCaption{
\begin{table}
\begin{center}
\begin{tabular}{|c|c|c|c||c|c|c|} \hline
& \multicolumn{3}{c||}{Konishi} & \multicolumn{3}{c|}{Spin 2}\\\hline
& $SU(2)$  			& $SU(3)$  & $SU(4)$  & $SU(2)$ 	& $SU(3)$  & $SU(4)$\\\hline
$\tau = i$ 			& $2.83\substack{+0.12\\-0.09}$ 	& $3.07\substack{+0.22\\-0.09}$ & $3.24\substack{+0.34\\-0.06}$ 
					& $5.16\substack{+0.16\\-0.14}$ 	& $5.50\substack{+0.28\\-0.15}$ & $5.73\substack{+0.46\\-0.11}$\\\hline
$\tau = e^{i\pi/3}$ & $2.76\substack{+0.12\\-0.17}$ 	& $2.98\substack{+0.20\\-0.18}$ & $3.15\substack{+0.29\\-0.19}$ 
					& $5.06\substack{+0.15\\-0.24}$ 	& $5.36\substack{+0.27\\-0.26}$ & $5.59\substack{+0.40\\-0.26}$\\\hline
Bound 				& $3.05$ 							& $3.38$ 						& $3.59$ 		
					& $5.32$ 							& $5.66$ 						& $5.80$ 						\\\hline
Corner				& $2.93$ 							& $3.24$	 					& $3.47$ 		
					& $5.28$ 							& $5.60$						& $5.75$ 						\\\hline
\hline
\end{tabular}
\end{center}
\end{table}}

\def\spinZeroTwoTable{
\begin{table}
\begin{center}
\begin{tabular}{|c|c|c|c||c|c|c|} \hline
& \multicolumn{3}{c||}{Konishi} & \multicolumn{3}{c|}{Spin 2}\\\hline
& $SU(2)$  			& $SU(3)$  & $SU(4)$  & $SU(2)$ 	& $SU(3)$  & $SU(4)$\\\hline
$\tau = i$ 			& $2.83\substack{+0.12\\-0.09}$ 	& $3.07\substack{+0.22\\-0.09}$ & $3.24\substack{+0.34\\-0.06}$ 
					& $5.16\substack{+0.16\\-0.14}$ 	& $5.50\substack{+0.28\\-0.15}$ & $5.73\substack{+0.46\\-0.11}$\\\hline
$\tau = e^{i\pi/3}$ & $2.76\substack{+0.12\\-0.17}$ 	& $2.98\substack{+0.20\\-0.18}$ & $3.15\substack{+0.29\\-0.19}$ 
					& $5.06\substack{+0.15\\-0.24}$ 	& $5.36\substack{+0.27\\-0.26}$ & $5.59\substack{+0.40\\-0.26}$\\\hline
Bound 				& $3.05$ 							& $3.38$ 						& $3.59$ 		
					& $5.32$ 							& $5.66$ 						& $5.80$ 						\\\hline
Corner				& $2.93$ 							& $3.24$ 						& $3.47$ 		
					& $5.28$ 							& $5.60$ 						& $5.75$ 						\\\hline
\hline
\end{tabular}
\end{center}
\caption{Interpolated values for spin zero and spin two operators at $\tau=i$ and $\tau=\exp(i \pi/3)$, along with the bounds and estimates for the same operators obtained from the conformal bootstrap.\label{t1}}
\end{table}}

\def\spinFourTable{
\begin{table}
\begin{center}
\begin{tabular}{|c|c|c|c|} 	\hline
					\multicolumn{4}{|c|}{Spin 4}								\\\hline
					& $SU(2)$ 			& $SU(3)$			& $SU(4)$ 			\\\hline
$\tau =i$ 			& $7.20-7.55$		& $7.59-8.09$ 		& $7.89-8.57$		\\\hline
$\tau = e^{i\pi/3}$ & $6.96-7.43$ 		& $7.29-7.92$  		& $7.56-8.34$ 		\\\hline
Bound 				& $7.55$ 			& $7.80$ 			& $7.89$ 			\\\hline
Corner 				& $7.53$ 			& $7.79$ 			& $7.88$ 			\\\hline
\hline
\end{tabular}
\end{center}
\caption{Interpolated values for the spin four operator at $\tau=i$ and $\tau=\exp(i \pi/3)$. The numbers shown represent the mean of the two-loop (lower) and the mean of the three-loop (upper) resummations, along with the bounds and estimates for the same operator obtained from the conformal bootstrap.\label{t2}}
\end{table}}